\documentclass[prl,twocolumn,showpacs,superscriptaddress,groupedaddress]{revtex4}

\usepackage{amssymb,amsfonts,amsmath} 

\usepackage{graphicx}

\newcommand{\bb}[1]{\mathbf{#1}}     % bold matrices 

\newcommand{\ket}[1]{\left | #1 \right\rangle} 
\newcommand{\bra}[1]{\left\langle #1 \right |}

\newcommand{\rr}{\bb{r}} 
 
\newcommand{\drr}{d\,\rr} 
\newcommand{\psid}{\psi^\dagger} 
\newcommand{\pp}{\bb{p}} 
\newcommand{\qq}{\bb{q}} 
 
\newcommand{\half}{\frac{1}{2}}

\begin{document}

%\preprint{}

\title{Spin-asymmetric Josephson effect} 


%\author{}
%\thanks{}
%\affiliation{}

\author{M.O.J.\ Heikkinen} 
\affiliation{Department of Applied Physics,  
Aalto University School of Science and Technology, P.O.Box 15100, FI-00076 Aalto, FINLAND} 
\author{F.\ Massel}
\affiliation{Department of Applied Physics,  
Aalto University School of Science and Technology, P.O.Box 15100, FI-00076 Aalto, FINLAND} 
\author{J.\ Kajala}
\affiliation{Department of Applied Physics,  
Aalto University School of Science and Technology, P.O.Box 15100, FI-00076 Aalto, FINLAND} 
\author{M.J.\ Leskinen}
\affiliation{Department of Applied Physics,  
Aalto University School of Science and Technology, P.O.Box 15100, FI-00076 Aalto, FINLAND} 
\author{G.S.\ Paraoanu}
\affiliation{Low Temperature Laboratory,  
Aalto University School of Science and Technology, P.O.Box 15100, FI-00076 Aalto, FINLAND} 
\author{P. T\"orm\"a}
\email{paivi.torma@hut.fi}
\affiliation{Department of Applied Physics,  
Aalto University School of Science and Technology, P.O.Box 15100, FI-00076 Aalto, FINLAND}


%\date{November 3, 2010}

\begin{abstract}
We propose that 
with ultracold Fermi gases one can realize a spin-asymmetric 
Josephson effect in which the two spin components of a Cooper pair 
are driven asymmetrically --- corresponding to driving a Josephson 
junction of two superconductors with different voltages 
$V_\uparrow$ and $V_\downarrow$ for spin up and down electrons, 
respectively.  We predict that the spin up and down components 
oscillate at the same frequency but with different amplitudes. 
Furthermore our results reveal that the standard interpretation of the 
Josephson supercurrent in terms of coherent bosonic pair tunneling is 
insufficient.  We provide an intuitive interpretation of the 
Josephson supercurrent as interference in Rabi oscillations of pairs and 
single particles, the latter causing the asymmetry. 
\end{abstract}

%\keywords{}

\maketitle

%\section{}
%\subsection{}
%\subsubsection{}

When a coherent many-body system is partitioned into two sub-systems,   
the dynamics of macroscopic observables such as relative number of   
particles and relative phase is called the Josephson effect   
\cite{Josephson1962a,Leggett2001a}. The   
external Josephson effect has been realized in superconducting junctions  
\cite{Anderson1963a,Langenberg1965a_etal},   
superfluid $^3$He \cite{Backhaus1997a_etal}  
and $^4$He \cite{Sukhatme2001a_etal}, and in 
Bose-Einstein condensates (BEC) of alkali atomic gases in double-well   
traps \cite{Albiez2005a_etal,Levy2007a_etal}. The internal Josephson effect has been  
demonstrated in $^3$He \cite{Webb1975a} 
and is expected to occur 
in spin BECs   
\cite{Leggett2001a,Hall1998b_etal,Bohi2009a_etal,Chang2005a_etal}. Also in the context of  
ultracold Fermi   
gases \cite{Giorgini2008a} 
the possibility of the Josephson effect has recently received theoretical 
interest \cite{Paraoanu2002a, 
Onofrio2004a,Salasnich2008a,Galitski2005b}. 
In this letter we show that partitioning a system of Cooper-paired   
fermions so that the two components of the pair experience different   
potentials (this is what we mean by ``spin-asymmetric'' here) 
leads to a novel effect, namely different-amplitude but   
phase-synchronized number-oscillations of the components.
Although the microscopic description of the Josephson effect is 
based on single particle tunneling, the standard interpretation 
of the Josephson supercurrent is given in terms of coherent 
tunneling of bosons or Cooper pairs \cite{deGennes1999a}. 
Importantly, our results 
show that such an interpretation is insufficient. We provide a clear, 
intuitive explanation of the predicted spin-asymmetric Josephson 
effect and a new understanding of the Josephson supercurrent as a 
process where not only pairs but also the spin-components separately 
contribute via interference.

We propose that the spin-asymmetric Josephson effect can be realized 
in a four-component Fermi gas in which two superfluids are coupled by 
radio-frequency (rf) fields (Fig.~1a), as in rf spectroscopy 
\cite{Regal2003b}. 
The setup is motivated by the recent realization of three-component 
Fermi gases \cite{Wenz2009a_etal}. 
Another possible, perhaps experimentally simpler, realization is a 
superfluid two-component ultracold Fermi gas (Fig. 1b) in a 
(spin)component-dependent double-well potential.  
The theoretical 
descriptions of the systems of Fig.~1a and Fig.~1b 
are identical (in this letter we use the notation 
of the former). 
We also suggest that the spin-asymmetric effect can be realized in a
SIS-junction of two materials with different Zeeman splittings 
\cite{Meservey1994a}.
 
The setup of Fig. 1a corresponds to a many-body Hamiltonian 
$H=H_0+H_{rf}$,
where 
$H_0 =\int\drr \sum_{i}\psid_i(\rr{})( -\frac{\nabla^2}{2m} -\mu_i )\psi^{}_i(\rr{}) 
+\half\int\drr\sum_{i\ne j} U_{ij}\psid_i(\rr{})\psid_j(\rr{})\psi^{}_j(\rr{})\psi^{}_i(\rr{})$.
Here $\psi^{}_i(\rr{})$ and $\psid_i(\rr{})$ are the fermionic 
field operators for the internal state $i$ and 
$\mu_i$ is the chemical potential (we assume $\mu_i\equiv\mu$),   
and $U_{ij}$ give the interaction strengths in the s-wave 
contact potential approximation. We assume that $U_{12}$ and $U_{34}$ 
lead to pairing and set $\hbar=1$. 

In the rotating wave approximation \cite{Cohen1998a} the tunneling 
coupling between states $\ket{i}$ and $\ket{j}$ is given by the
Rabi frequency $\Omega_{ij}$ with the detuning 
$\delta_{ij}=\nu_{ij}-\omega_{ij}$.  Here $\nu_{ij}$ is the frequency 
of the field and $\omega_{ij}$ is the resonance frequency of the 
hyperfine transition.  The effect of the electromagnetic field on the 
system is then described by 
$
H_{rf}= 
\frac{\delta_{13}}{2} \int\drr  
(\psid_1(\rr{})\psi^{}_1(\rr{})-\psid_3(\rr{})\psi^{}_3(\rr{}))
+\frac{\delta_{24}}{2}\int\drr  
(\psid_2(\rr{})\psi^{}_2(\rr{})-\psid_4(\rr{})\psi^{}_4(\rr{}))
+\Omega_{13}\int\drr \, 
\psid_1(\rr{})\psi^{}_3(\rr{}) + \textit{h.c.} 
+\Omega_{24}\int\drr \, 
\psid_2(\rr{})\psi^{}_4(\rr{}) + \textit{h.c.}
%H_{rf}= 
%\frac{\delta_{13}}{2} \int\drr  
%\vas\psid_1(\rr{})\psi^{}_1(\rr{})-\psid_3(\rr{})\psi^{}_3(\rr{})\oik
%+\frac{\delta_{24}}{2}\int\drr  
%\vas\psid_2(\rr{})\psi^{}_2(\rr{})-\psid_4(\rr{})\psi^{}_4(\rr{})\oik
%+\Omega_{13}\int\drr \, 
%\psid_1(\rr{})\psi^{}_3(\rr{}) + \textit{h.c.} 
%+\Omega_{24}\int\drr \, 
%\psid_2(\rr{})\psi^{}_4(\rr{}) + \textit{h.c.}
$    
In analogy to the usual Josephson junctions, the states $\ket{1}$ and 
$\ket{2}$ correspond to spin up and down electrons in the 
left side superconductor, $\ket{3}$ and $\ket{4}$ on the right.  
The detunings $\delta_{13}$ and $\delta_{24}$ play the role of the voltage. 

The essential new feature in atomic gases is the possibility to 
set $\delta_{13} \neq \delta_{24}$, 
which in the case of the Josephson junction corresponds 
to spin-dependent voltages. Note that this is different
from superconductor-ferromagnet-superconductor (SFS) 
structures \cite{Bergeret2005}
in which the spin-active barrier coupling plays the crucial role.
Though in our case also the couplings could be different, only
the spin-asymmetric potential is relevant for our predictions. 
Moreover, one might consider ferromagnetic superconductors
\cite{Saxena2000} in a junction
as a related system but those materials have most likely an exotic ground
state which does not fit our description.

We now determine the transition rates (\textit{i.e.} 
particle currents) between states $\ket{1}$ and $\ket{3}$, 
$I_{13}(t)\equiv\langle \dot{N}_1\rangle$, and between states 
$\ket{2}$ and $\ket{4}$, $I_{24}(t)\equiv\langle \dot{N}_2\rangle$. 
We calculate a self-consistent linear 
response with respect to $H_{rf}$  
(valid when the number of transferred particles is small 
compared to the total particle number)  
for the system of Fig. 1 
with the aid of the Kubo formula and the Kadanoff-Baym method 
\cite{Baym1961a}. We obtain the currents
\begin{align} 
I_{13}(t)&=I^{S}_{13}+I^{C}_{13} \sin[(\delta_{13} + \delta_{24})t+\varphi], \notag \nonumber \\ 
I_{24}(t)&=I^{S}_{24}+I^{C}_{24} \sin[(\delta_{13} + \delta_{24})t+\varphi]\label{eq:currents}. 
\end{align} 
Here $I^S$ is the standard single particle (quasiparticle) 
current that occurs only for detunings $\delta_{ij}$ above the 
excitation gap $2\Delta$.  
The initial phase of the Josephson current is $\varphi$.  
The critical Josephson currents $I^C$ become, in a spatially 
homogeneous case and in the BCS description, 
\begin{align}  
I^{C}_{13} = &2 \left|\Omega_{13}\Omega_{24} 
\Pi_{\mathcal{F}}(\pp=0 , \delta_{24}+i\eta^+)\right| 
, \label{eq:KBjose1}\\ 
I^{C}_{24} = &2 \left|\Omega_{13}\Omega_{24} 
\Pi_{\mathcal{F}}(\pp=0 , \delta_{13}+i\eta^+)\right| 
, \label{eq:KBjose2} 
\end{align} 
where
$
\Pi_{\mathcal{F}} (\pp,\omega)=  
\frac{1}{\beta V}\sum_{\qq,\chi}  
\mathcal{F}_{12} (\qq,\chi)  
\mathcal{F}_{34}^* (\qq-\pp,\chi-\omega). 
$
Here $V$ is the volume and $\beta = 1/(k_B T)$ ($T$ is temperature and 
$k_B$ the Boltzmann constant). 
$\mathcal{F}_{12}=\mathcal{F}_{21}^*$ ($\mathcal{F}_{34}=\mathcal{F}_{43}^*$)  
is the anomalous mean field Matsubara Green's function for the superfluid of 
components $\ket{1}$ and $\ket{2}$ ($\ket{3}$ and $\ket{4}$).  
For details see supplementary material \cite{supp_josephson}. 

The striking result is that the critical current $I^{C}_{13}$ in 
equation \eqref{eq:KBjose1} depends only on
$\delta_{24}$, and similarly $I^{C}_{24}$ in equation 
\eqref{eq:KBjose2} only on $\delta_{13}$ (this is 
not limited to the BCS regime; it remains true whenever pairing 
correlations exist). By choosing different detunings $\delta_{13}$ and 
$\delta_{24}$, one can observe a spin-asymmetric Josephson effect in 
which the currents in the two tunneling channels are different in 
amplitude, but oscillate at the same frequency. 
Moreover, the results predict a tunable DC Josephson effect: by 
choosing the detunings so that $\delta_{13}+\delta_{24}=0$, 
the phase factor in equation (\ref{eq:currents}) is
constant but the critical current can still be tuned.
The conclusions hold for experimentally realistic parameters and 
also if cross interactions (see Fig.1a) 
and finite temperature are included into our analysis, as shown in  
the supplementary material \cite{supp_josephson}. 
For typical parameters and taking \textit{e.g.} $\delta_{13}/E_F=0.4$ and $\delta_{24}/E_F=0.5$, with $E_F$
the Fermi energy, one obtains a considerable asymmetry of $I^C_{13}/I^C_{24}=1.14$, see \cite{supp_josephson} for details. 
By performing the self-consistent calculation  
we have removed the 
ambiguity of whether our previous suggestion of the critical current 
asymmetry \cite{Paraoanu2002a} was due to a simple linear 
response approach following the Ambegaokar-Baratoff treatment 
\cite{Ambegaokar1963a}.   

\begin{figure}
\includegraphics[width=0.44\textwidth]{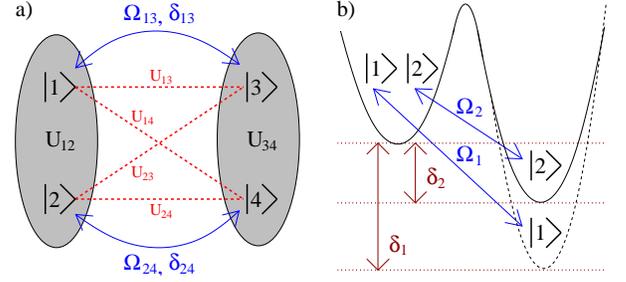} 
\caption 
{ 
Spin-asymmetric Josephson effect setup.
(a) The four-component 
Fermi gas. Particles in internal (\textit{e.g.} hyperfine) states 
$\ket{1}$ -- $\ket{4}$ coexist spatially.
The components $\ket{1}$ and $\ket{2}$ as well as $\ket{3}$ and 
$\ket{4}$ form Cooper pairs due to the interactions $U_{12}$ and 
$U_{34}$.  The cross-interactions $U_{13}$, $U_{14}$, $U_{23}$, 
$U_{24}$ are assumed weak, thus not leading to pairing.  The 
tunneling is driven by applying rf fields to couple the states.  The 
Rabi couplings $\Omega_{ij}$ correspond to the weak tunneling link 
between the two superconductors in a Josephson junction.  The 
detunings $\delta_{ij}=\nu_{ij}-\omega_{ij}$ play the role of the 
voltage, $eV$.  Remarkably, one can choose $\delta_{13} \neq 
\delta_{24}$ and create the analogue of a spin-dependent voltage in a 
Josephson junction.  The states $\ket{1}$ and $\ket{3}$ ($\ket{2}$ and 
$\ket{4}$) must be states of the same atom to allow the rf coupling, 
but $\ket{1}$ and $\ket{2}$ can be \textit{e.g.}
$^6$Li and $^{40}$K.  (b) Cooper pairs of a two-component Fermi 
gas in a spin dependent double-well potential.  This setup is 
conceptually equivalent to the system of (a) albeit the absence 
of cross-interactions. 
} 
 \label{fig:system} 
 \end{figure}  
 
Our results \eqref{eq:currents}-\eqref{eq:KBjose2} are in obvious contradiction 
with the standard interpretation of the Josephson supercurrent in terms of 
coherent \emph{pair} tunneling excluding any difference in 
the Josephson currents of different spin-components. 
In what follows we provide an explanation  
for the spin-asymmetric Josephson effect 
which not only resolves this paradox but also opens 
up a new point of view to the conventional Josephson effect.  
Let us begin by writing down  
the initial state of the two superfluids 
as a product state of two BCS states: 
$
\ket{\Psi} = \ket{\rm BCS}_{12}\otimes\ket{\rm BCS}_{34} 
           =  \prod_k \left(u_k \ket{0}_k + v_k\ket{12}_k\right)  \prod_{k'} \left(u_{k'} 
\ket{0}_{k'} + v_{k'} \ket{34}_{k'}\right) 
$
where $\ket{ij}_{k} = c^{\dagger}_{k,i}c^{\dagger}_{-k,j}\ket{0}_{k}$.
Next, we single out one Cooper pair from each superfluid.
Since the rf coupling between 1-3 and 2-4 conserves momentum we focus on states with $k=k'$ (the momentum conservation can 
be relaxed and our conclusions still hold): 
\begin{align} 
& (u_k \ket{0}_k + v_k \ket{12}_k)  (u_k \ket{0}_k + v_k \ket{34}_k) 
 = u_k^2 \ket{0}_k\ket{0}_k \nonumber \\ 
& + v_k^2 \ket{12}_k \ket{34}_k  + u_k v_k\ket{12}_k \ket{0}_k   
 + u_k v_k \ket{0}_k \ket{34}_k. \label{buildingblock} 
\end{align} 
The empty state $\ket{0}\ket{0}$ cannot contribute to the current 
and neither can $\ket{12} \ket{34}$ since it is Pauli blocked. 
Therefore, the Josephson physics arises from the $u_k v_k 
\ket{12} \ket{0} + u_k v_k \ket{0}\ket{34}$ superposition. 
 
We then ask whether the essential features of our results can be  
explained by considering the dynamics of a single Cooper pair, initially 
in the above superposition state characteristic for the BCS state with a macroscopic phase. 
In additition to the paired states 
$\ket{12}\ket{0}\equiv\ket{12}$ and $\ket{0}\ket{34}\equiv\ket{34}$
the states $\ket{1}\ket{4}\equiv\ket{14}$ and $\ket{2}\ket{3}\equiv\ket{23}$ 
are required to have a closed subsystem with respect to the tunneling coupling, see Fig. 2. 
These broken pair states are analogous to single particle excitations within the BCS formalism. 
 
We now solve the time evolution of this system perturbatively 
in the couplings $\Omega_{ij}$. 
We consider an initial state in the general superposition form  
$\alpha_{\rm I} \ket{12}+\beta_{\rm II}\ket{34}$ as suggested by equation \eqref{buildingblock}.  
The total current becomes 
\begin{align} 
\label{eq:nump_1} 
% \bra{\phi(t)}\dot{N}_1\ket{\phi(t)}= 
%     & |\alpha_{\rm I}|^2 \bra{\phi_{\rm I}(t)}\dot{N}_1\ket{\phi_{\rm I}(t)}          \nonumber\\ 
%    +& |\beta_{\rm II}|^2 \bra{\phi_{\rm II}(t)}\dot{N}_1\ket{\phi_{\rm II}(t)}    \nonumber\\ 
%    +& \alpha_{\rm I} \beta_{\rm II}^*\bra{\phi_{\rm II}(t)}\dot{N}_1\ket{\phi_{\rm I}(t)} \nonumber\\  
%    +& \alpha_{\rm I}^* \beta_{\rm II}\bra{\phi_{\rm I}(t)}\dot{N}_1\ket{\phi_{\rm II}(t)}, 
&\bra{\phi(t)}\dot{N}_1\ket{\phi(t)} = |\alpha_{\rm I}|^2 \bra{\phi_{\rm I}(t)}\dot{N}_1\ket{\phi_{\rm I}(t)} \nonumber\\          
    &+ |\beta_{\rm II}|^2 \bra{\phi_{\rm II}(t)}\dot{N}_1\ket{\phi_{\rm II}(t)}    
    + \alpha_{\rm I} \beta_{\rm II}^*\bra{\phi_{\rm II}(t)}\dot{N}_1\ket{\phi_{\rm I}(t)} \nonumber\\ 
    &+ \alpha_{\rm I}^* \beta_{\rm II}\bra{\phi_{\rm I}(t)}\dot{N}_1\ket{\phi_{\rm II}(t)}, 
\end{align} 
where $\ket{\phi_{\rm I}(t)}=\exp(-i H t)\ket{12}$ and 
$\ket{\phi_{\rm II}(t)}=\exp(-i H t)\ket{34}$ are calculated 
to second order in $\Omega_{ij}$.
For details and for a complementary discussion in terms of exact numerical eigenstates 
(dressed states) see \cite{supp_josephson}.  
Here the first two terms contain only single particle Rabi processes,   
which correspond to the standard single 
particle (quasiparticle) currents in a Josephson junction, $I_{ij}^S$ 
in equation  \eqref{eq:currents}.  Only the last two terms in equation 
\eqref{eq:nump_1}, with $\alpha_{\rm I} 
\beta_{\rm II}^*\equiv|\alpha_{\rm I} \beta_{\rm II}|e^{i\varphi}$, 
contribute to the Josephson current. 
Thus, the Josephson effect originates from the interference 
  part of the Rabi oscillations in the $\ket{12}$, $\ket{34}$, 
  $\ket{14}$, $\ket{23}$ state space.

Isolating the terms oscillating at the Josephson frequency in 
$\bra{\phi(t)}\dot{N}_1\ket{\phi(t)}$, we get the Josephson current 
\begin{align} 
  \label{eq:NJoseph} 
   I_{13}^J&= I^C_{13}(\delta_{24}) \sin[(\delta_{13}+\delta_{24})t+\varphi],\nonumber\\  
I^C_{13}(\delta_{24}) &=2\Omega_{13}\Omega_{24}|\alpha_{\rm I}\beta_{\rm II}| \left[ \frac{1}{U+\delta_{24}} + \frac{1}{U-\delta_{24}} \right].
\end{align} 
Hereby, we obtain the same qualitative result as our linear response 
calculation gave in equation \eqref{eq:currents}: the 
amplitude of $I_{13}^J$ depends only on the detuning $\delta_{24}$. 
Now we can identify the source of the dependence of $I^C_{13}$ on $\delta_{24}$. 
The current $I^J_{13}$ derives from the population of species $\ket{1}$, 
\emph{i.e.} both the population of state $\ket{12}$ and $\ket{14}$. 
The result of equation \eqref{eq:NJoseph} is the sum of these two
contributions.

The contribution from state $\ket{12}$
is the result of tunneling between the paired states $\ket{12}$ and 
$\ket{34}$, via the intermediate state $\ket{14}$ (or $\ket{23}$)
as shown in Fig. 3a. 
The term is symmetric in $\delta_{13}$ and $\delta_{24}$ and proportional to 
$
M_{pair} = \frac{1}{U+\delta_{13}}+ 
    \frac{1}{U+\delta_{24}}. 
$
The contribution is of second order because it originates from a 
zeroth order and a second order process: the population due to pair 
tunneling from $\ket{34}$ to $\ket{12}$ (second order) is 
indistinguishable from the initial (zeroth order) population in state 
$\ket{12}$. Pair tunneling contributions that do not originate from interference
also exist but they are of fourth order in $\Omega$. 
 
The contribution to $I_{13}^C$ from state $\ket{14}$ arises from interference of
broken pairs (single particles) as depicted in Fig. 3b.  
This contribution is responsible for the asymmetry
as it is proportional to 
$
M_{single} = \frac{1}{U-\delta_{24}}-\frac{1}{U+\delta_{13}}. 
$
Physically, this term arises because we cannot distinguish between 
broken pairs from the states $\ket{12}$ and $\ket{34}$.
The two paths leading to the state 
$\ket{14}$ are not symmetric with respect to  
$\delta_{13}$ and $\delta_{24}$, see Figs. 2 and 3.
This gives important new insight also to the Josephson effect in general:  
also \textit{single particle} interferences 
contribute to the Josephson current
$I_{ij}^C \sin (\delta_{13} + \delta_{24})$ of equation \eqref{eq:currents}.
 
In the standard symmetric case  
$\delta_{13}=\delta_{24}\equiv\delta$, we have $M_{pair} = 2/(U+\delta)$ 
and $M_{single}=2\delta/(U^2-\delta^2)$.
At the DC Josephson limit, $\delta\rightarrow 0$, 
the pair transfer $M_{pair}$ dominates, 
which is intuitively appealing. 
Closer to the Riedel peak the excited state (single  
particle) interference $M_{single}$ becomes equally important.
To illustrate further, in a typical Al/AlO$_x$/Al junction 
at the voltage of $0.015~\mathrm{mV}$ 
the single particle interference accounts for $1.7~\%$ of the AC Josephson current. 
Note that our "single particle interference term" is not the cosine-term 
(also called "quasiparticle interference term" \cite{Waldram1976}) 
of the Josephson effect. The cosine-term involves real single 
particle transitions and exists only for voltages above $2\Delta$ 
at zero temperature. In contrast, our single particle interference term 
is inherent in the supercurrent and corresponds to virtual transitions. 
Similarly, our findings are different from various combined effects of 
single particle currents and supercurrents in small Josephson junctions 
\cite{Fulton1989}, 
where again the single particle transitions are real, not 
virtual. Note also that we do not consider any interactions between 
the Cooper pairs (analogue of charging effects) nor the effect of 
the environment.
 
\begin{figure}
\includegraphics[width=0.34\textwidth]{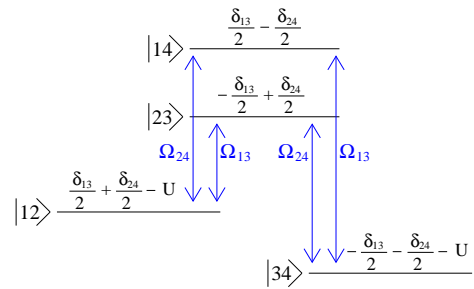} 
\caption 
{ 
Energy level diagram of the four-state model.   
The basis states $\ket{12}$, $\ket{34}$, $\ket{14}$, $\ket{23}$ form a 
closed subspace under the Rabi oscillation processes caused by the rf coupling.
The paired states $\ket{12}$ and $\ket{34}$ have an energy lower by $U$ than the states of 
broken pairs $\ket{14}$ and $\ket{23}$. 
} 
\label{fig:level} 
\end{figure} 
 
\begin{figure}
\includegraphics[width=0.22\textwidth]{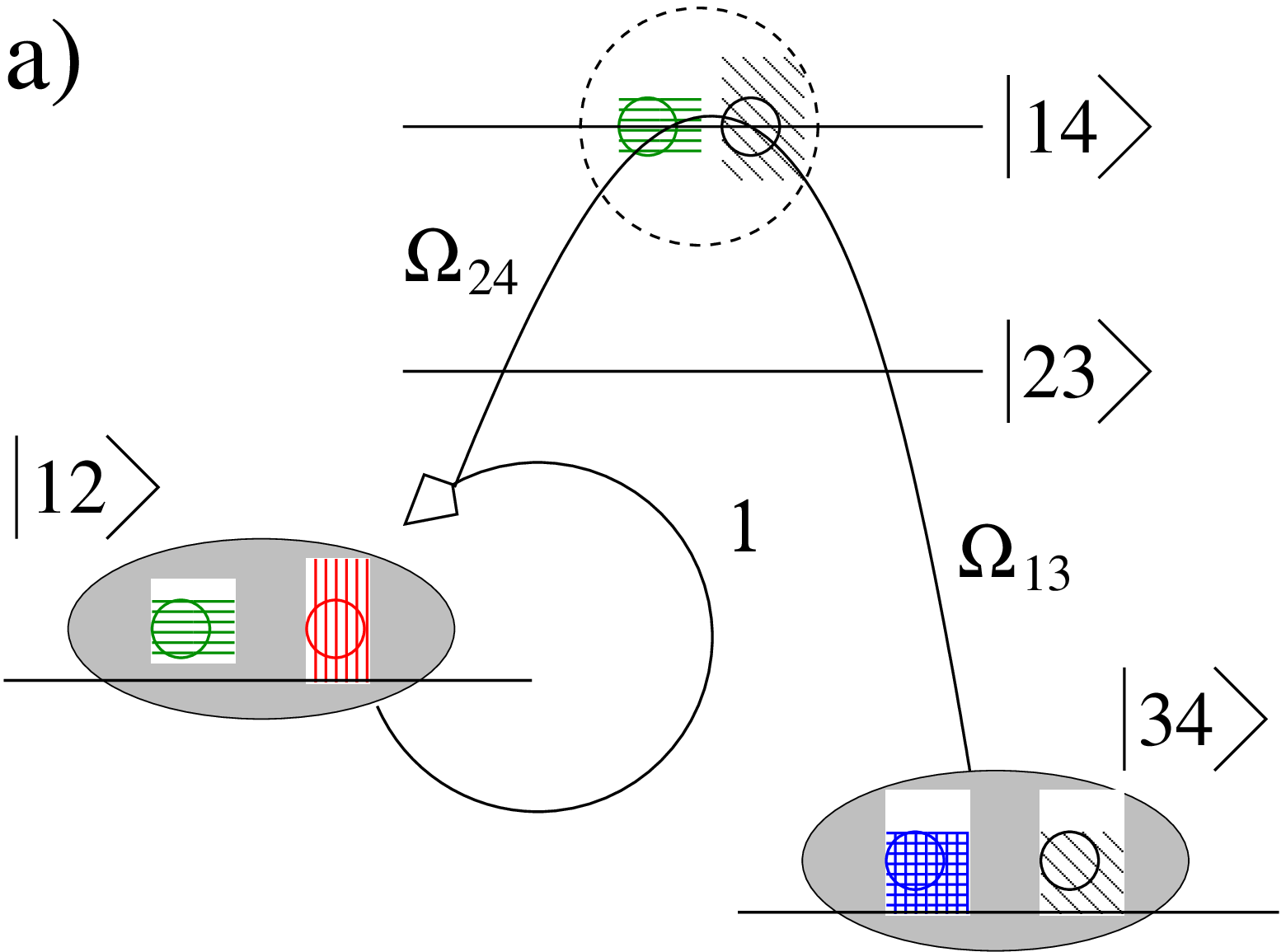} 
\includegraphics[width=0.22\textwidth]{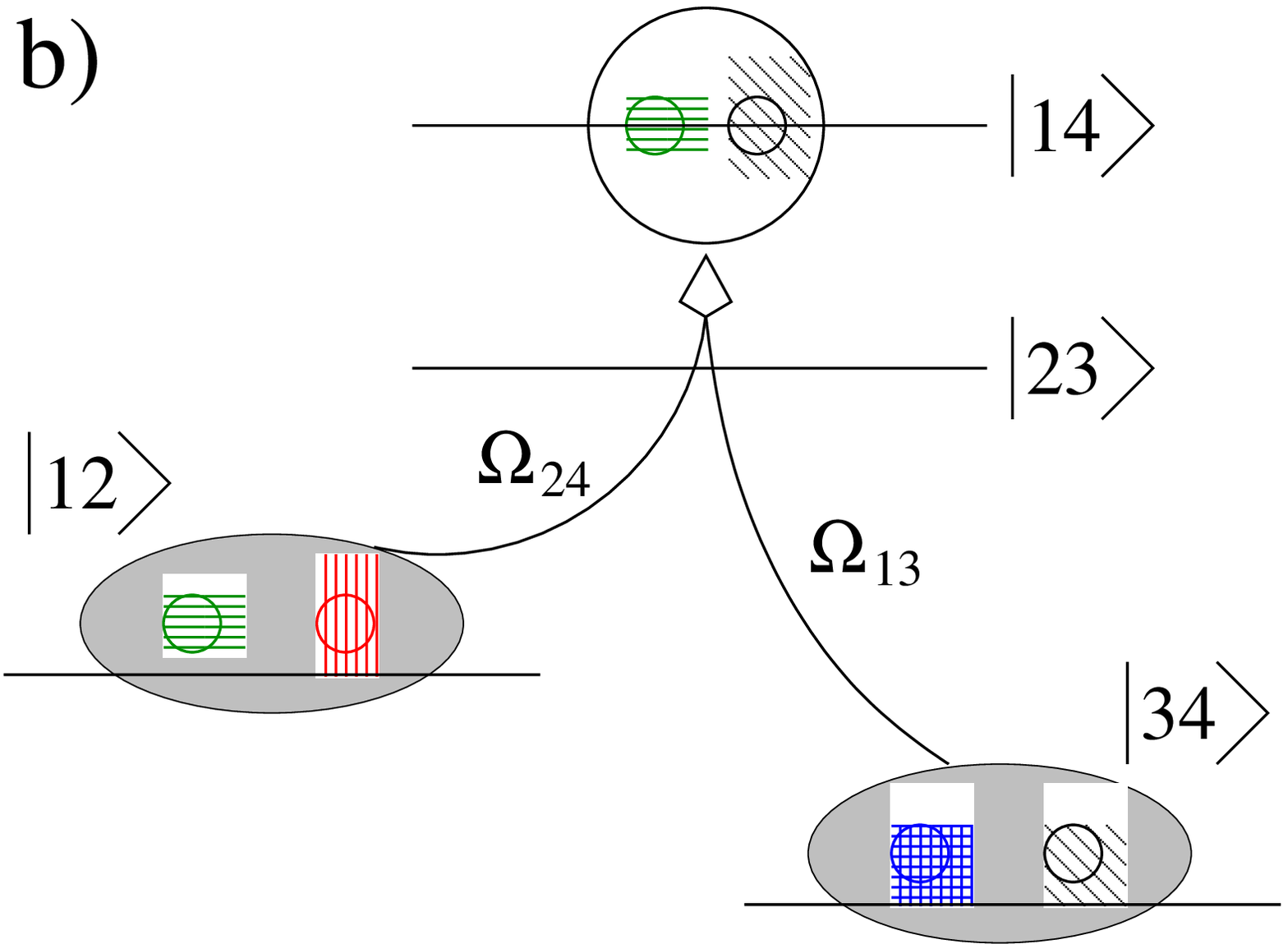} 
\caption 
{ 
Interference processes leading to the Josephson supercurrent.  The 
initial state is $\ket{\phi}=(\ket{12}+\ket{34})/\sqrt{2}$.  (a) The 
second order transition from $\ket{34}$ through $\ket{14}$ to 
$\ket{12}$ creates an interference with the initial (zeroth order) population in 
state $\ket{12}$.
The resulting contribution is of the order 
$\Omega_{13}\Omega_{24}$.   
Also the same process via the state $\ket{23}$ contributes.  
(b) The first order transitions to the excited state 
$\ket{14}$ from $\ket{12}$ and $\ket{34}$ create an interference 
proportional to $\Omega_{13}\Omega_{24}$. 
} 
\label{fig:pair_interference} 
\end{figure}  
 
We emphasise the interference nature of the Josephson effect: it 
requires lack of which-way information on the tunneling path of the 
particle. Above, the superposition $u_k v_k (\ket{12}\ket{0} + 
\ket{0}\ket{34})$ was due to the uncertainty of the particle number in 
the BCS state.  It is interesting to contrast this with the 
number-projected BCS-state \cite{Bayman1960a} or the Fock-state 
\cite{Leggett}, 
$ 
\ket{\Psi}_{Ncons} =
(u_{k} v_{k'} \ket{0}_{k} \ket{12}_{k'}  + u_{k'} 
v_{k} \ket{12}_{k}  \ket{0}_{k'} ) 
(u_{k} v_{k'} \ket{0}_{k} \ket{34}_{k'}  + u_{k'} v_{k}\ket{0}_{k'} \ket{34}_{k})
$
(for two $k$-states.)
Here the part relevant for
Josephson physics would be 
$
u_{k} v_{k'} u_{k'} v_{k} \ket{12}_{k'}  \ket{34}_{k}  
+ u_{k} v_{k'} u_{k'} v_{k} \ket{12}_{k} \ket{34}_{k'}. 
$
Now, the entanglement of $\ket{34}_{k'}$ with 
$\ket{12}_{k}$ allows to determine, whether a particle 
in state $\ket{1}$ belongs to an initial 1-2 pair or to a tunnelled 
3-4 pair. The indistinguishability is lost and there is no Josephson 
effect. 
Note that the system of Fig. 1a should be cooled with the rf couplings on
to realize the necessary uncertainty in particle number.
One can also ask whether, in 
the case of the separated 1-2 and 3-4 condensates (Fock 
states), the measurement process itself is sufficient to generate the 
relative particle number uncertainty, 
in analogy to the famous problem of interference between 
two BECs \cite{Andrews1997a_etal}.
 
Importantly, based on the picture of 
interfering Rabi processes, one would 
anticipate corrections from the coupling strength $\Omega_{ij}$ to the Josephson 
frequency $\delta_{13}+\delta_{24}$ given by linear response.
We have observed such a correction in the numerical simulations, see supplementary material \cite{supp_josephson}.
We expect this correction 
to vanish in the thermodynamic limit,
but it could be experimentally tested in ultracold gases and in superconducting grains. 
 
In summary, we have predicted a spin-asymmetric Josephson effect which 
could be observed in ultracold Fermi gases and solid state systems and is 
likely to be relevant for spintronics. We provide a microscopic 
description of the Josephson effect as interference in Rabi 
oscillations of pairs and single particles, where the latter cause the 
predicted asymmetry.  
In particular, our finding that the interference part 
of single particle currents contributes to the Josephson supercurrent, in addition to the 
interference in pair tunneling, is fundamental. The single particle interference 
contribution is present already in the standard (symmetric) case, and becomes manifest 
and experimentally verifiable in the asymmetric one.

\begin{acknowledgments} 
We thank J.J.\ Kinnunen and S.\ van Dijken for useful discussions. 
This work was supported by NGSMP, FICS, EUROQUAM/FerMix and Academy 
of Finland (Project No. 129896, No. 135135, No. 210953, No. 213362, No. 217043, No. 217045),
and conducted as a part of a EURYI scheme grant \cite{ack_euryi}.  We acknowledge the use of CSC -- IT Center for 
Science Ltd computing resources. 
\end{acknowledgments} 

%\bibliography{jose_v4.bib}

\end{document}

{\centering\section*{Setup for the Josephson effect}}
 \label{sec:jos_eff}

 In this section we describe in greater detail the setup we
 considered in the linear response calculation, relaxing some of the
 conditions assumed in the article.
 Altough we use the notation and terminology of Figure 1a of the main article
 the description is equivalent for example with the system of Figure 1b
 of the main article.
 
 In order to accommodate a superfluid with balanced pairing, we require a
 balanced number of particles within the superfluid states \ie~$N_1=N_2$ and
 $N_3=N_4$. We further assume that for any tunneling in the system the
 relation ${\delta N_i}/{N_i}\ll 1$ holds, so that we do not have
 to take into account the internal dynamics of the superfluids. In
 particular, this assumption implies that the superfluid states are not
 destroyed because of the transitions. The tunneling link between the
 hyperfine states is provided by two radio frequency (RF) fields
 $\E_{13}(\rr,t)$ and $\E_{24}(\rr,t)$.  The field $\E_{ij}(\rr,t)$
 drives transitions between the states $\ket{i}$ and $\ket{j}$. For
 simplicity we take these fields as a single mode with frequency
 $\nu_{ij}$ and wave vector $\kk_{ij}$.  The strength of the coupling
 between the electric field $\E_{ij}(\rr,t)$ and the hyperfine
 transition between states $\ket{i}$ and $\ket{j}$ is given by the
 Rabi frequency \be \Omega_{ij} (\rr) = \dd_{ij}\cdot\E_{ij}(\rr),
 \label{eq:rabi}\ee where $\dd_{ij}$ is the electric dipole moment of
 the transition.  Throughout the treatment we assume that the RF
 fields are near resonance and that the intensity of the radiation is
 low. With these assumptions we may use the rotating wave
 approximation \cite{cohen_tann}.  As a result, it is possible to
 transform the time dependence of the field in the
 rotating frame into a shift in the transition
 frequency  $\delta_{ij}$, called detuning, defined as \be
 \delta_{ij} = \nu_{ij}-\omega_{ij}, \ee where $\omega_{ij}$ is the
 frequency of the transition.

 Within the same superfluid state the hyperfine states will naturally have
 the same chemical potential, therefore $\mu_1=\mu_2$ and $\mu_3=\mu_4$. In
 the article we further assumed $\mu_1=\mu_3$, but here we remove this
 constraint.  Including this difference in the chemical potential
 between the two superfluids into the RF detuning, we get the modified
 detuning of the hyperfine transition 
\be 
\tilde{\delta}_{ij} =\delta_{ij}+ \mu_j-\mu_i. \label{moddet} 
\ee 
This quantity describes
 the energy gain or loss associated with a particular hyperfine
 transition without many-body effects.
\begin{figure} [!h]
\begin{center}
 \epsfig{file=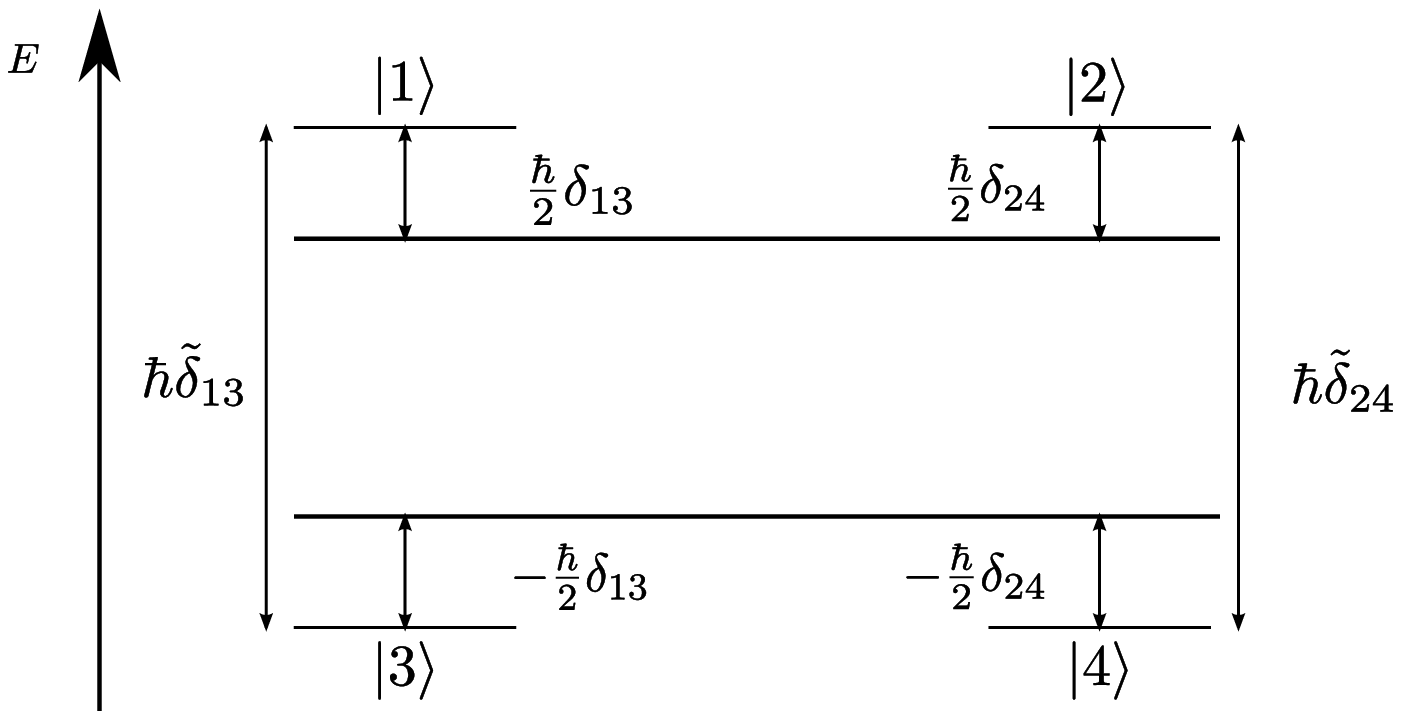,width=0.6\columnwidth}
\caption{The energy levels of the system.  In the unperturbed system
  the chemical potentials are equal within the BCS state:
  $\mu_1=\mu_2$ and $\mu_3=\mu_4$.  The detunings $\delta_{13}$ and
  $\delta_{24}$ of the RF fields can create an effective asymmetry in
  the energies.}
\label{fig:energiatasot}
\end{center}
\end{figure}

The energy levels of the system in question are shown in
Figure \ref{fig:energiatasot}.  Here we make use of the assumption that
the applied RF field does not alter the internal dynamics of the superfluids
so that we may split the detuning $\delta_{ij}$ between the
states $\ket{i}$ and $\ket{j}$ freely. We choose a symmetric
splitting, although we will eventually see that this choice does not
affect the results.

{\centering\section*{Linear response description}}
 \label{sec:lin_resp}

The system is described by the Hamiltonian
\begin{equation}
H=H_0+H_{RF}.
\end{equation}
Here we have defined $H_0$ as
\begin{align}
\label{eq:hzero}
H_0 =&\int\drr \sum_{i}\psid_i(\rr{})\left( -\frac{\nabla^2}{2m} -\mu_i \right)\psi^{}_i(\rr{}) \nonumber\\
+&\half\int\drr\sum_{i\ne j} U_{ij}\psid_i(\rr{})\psid_j(\rr{})\psi^{}_j(\rr{})\psi^{}_i(\rr{}).
\end{align}
The external perturbation is given by
 \be 
 H_{RF} = H_{c}+H_{d},
 \ee
 where we have
 \begin{eqnarray}
   \label{eq:Rf_split}
   H_{c} &=& \int\drr\sum_{i,j}
              \Omega_{ij}\psid_{i}(\rr{})\psi_{j}(\rr{}) \\
   H_{d}&=& \int\drr\sum_{i}
              W_{i}\psid_{i}(\rr{})\psi_{i}(\rr{}),
 \end{eqnarray}
 with the definitions $W_1=\mu_1+\delta_{13}/2$, $W_3=\mu_3-\delta_{13}/2$
and similarly for $W_2$ and $W_4$. The inclusion of the chemical potential
in these expressions is just a mathematical trick to aid the forthcoming calculation.
For RF fields it is a very good approximation to take the Rabi frequency $\Omega$ 
as independent of position.

The operator for current density to one hyperfine state is given by
the time derivative of the particle density operator, which
is obtained using the Heisenberg equation of motion
\be
\aikad{}N_{i}(\rry) 
= i \com{H}{N_{i}(\rry)}.
\ee
Only $H_c$ fails to commute with the number operator. 
Hence the Heisenberg equation of motion for the number operator
becomes
\begin{align}
\aikad{}N_{i}(\rry)  &= i \com{H_c(t)}{N_{i}(\rry)}\nonumber\\
&= i \intdrrsj{2} \Big(\Omega_{j\,i}\psid_{j}(\rrs{2})\psi_{i}(\rrs{1})%\nonumber\\
- \Omega_{i\,j}\psid_{i}(\rrs{1})\psi_{j}(\rrs{2})\Big). \label{heis}
\end{align}
The physical current density $  I_{i}(\rry) $
is the thermodynamic average of the current density operator
defined as $  I_{i}(\rry t) \equiv \aver{\dot{N}_{i}(\rry) }$,
which we then need to calculate.
However, the exact calculation of this quantity is not feasible,
since time evolution of the operators in the Heisenberg picture
depends on the coupling $H_c$. Hence, we need to resort to an
approximate solution with respect to the coupling strength. 
At this point it is most convenient to  
make a transformation to the interaction picture with respect to
$H_c$. For a general operator $A(t)$ this transformation is given by
\begin{equation}
A(t)=S(t,t_0)^{-1}A(t_0)S(t,t_0),
\end{equation}
in which $t_0$ gives the initial condition and
\begin{equation}
S(t,t_0)=T\exp\vas-i\tint{t_0}{t}H_c(\tpri)\oik.
\end{equation}
Assuming that $H_c$ is a weak perturbation, it is a good approximation 
to linearise this transformation with respect to $H_c$, in which case we have
\begin{equation}
S(t,t_0)=1-i\tint{t_0}{t}H_c(\tpri),
\end{equation}
and consequently $A(t)$ is linearised to
\begin{equation}
A(t)=A(t_0)-i\tint{t_0}{t}\com{A(t)}{H_c(\tpri)}.
\end{equation}
The thermodynamic expectation value of a generic operator $A(t)$ then follows the Kubo formula
\begin{equation}
\langle A(t) \rangle = \langle A(t_0) \rangle 
-i\tint{t_0}{t}\aver{\com{A(t)}{H_c(\tpri)}}.
\label{kubo_formula}
\end{equation}
Here $t_0$ is the reference time for the initial condition. 
Since we assume a system initially in equilibrium, there is no current
at the time $t_0$. Therefore, upon replacing $A$ with
the current operator $\dot{N}$, 
the first term in the expression (\ref{kubo_formula}) vanishes.
In order to obtain the Kubo formula in the frequency domain, 
it is a common strategy
to take the limit $t_0\rightarrow -\infty$ for the initial condition. 

Inserting equation (\ref{heis}) into equation (\ref{kubo_formula}) then yields the current density
\begin{align}
  I_{i}(\rry t) = &~i \tint{-\infty}{\infty}\intdrrsss{2}{3}{4} 
  \Big( \exp\left[{i\left(W_{l}-W_{i}\right)t+i\left(W_{j}-W_{k}\right)\tpri}\right]\nonumber\\
  &\times \Omega_{l i} \Omega_{j k}
  \AfunSp{1}{3}{4}{2}{t}{\tpri} - h.c. \Big),
  \label{eq:curr_gen}
\end{align}
where the retarded linear response function 
$\AfunSp{1}{2}{3}{4}{t}{\tpri} $ is defined as
\be
\AfunSp{1}{2}{3}{4}{t}{\tpri} 
= 
-i\theta(t-\tpri)\aver{ \com{ \psid_{k}(\rrs{3}t) \psi_{i}(\rrs{1}t)}{\psid_{l}(\rrs{4}\tpri)\psi_{j}(\rrs{2}\tpri)} }. 
\ee
In equation (\ref{eq:curr_gen}) we have used standard algebraic manipulations
to separate out the effect of the quantities $W_i$ from the remaining time
evolution of the field operators. 
For the hyperfine species $\left|1\right\rangle$, equation
\eqref{eq:curr_gen} can be rewritten as
\be
I_{13}(\rry t) = I^{S}_{13}(\rry t)+I^{J}_{13}(\rry t),  
\ee
where the standard single particle (quasiparticle) 
current $I^{S}_{13}$ is defined as
\be
 I^{S}_{13}(\rry t)= -2\intdrr_{2} 
\im\left[\Omega_{13}^*\Omega_{13} L_{1331}(\rry - \rrk ,\deti_{13}+i\eta^+)\right], \label{Is}
\ee
and the Josephson current $I^{J}_{13}$ as
\begin{align}
 I^{J}_{13}(\rry t)= -2\intdrr_{2} \im\Big[&e^{-i(\deti_{13} + \deti_{24})t}\Omega_{13}^*\Omega_{24}^*\nonumber\\ 
&\times L_{1234}(\rry -\rrk , -\deti_{24}+i\eta^+)\Big]. \label{Ij}
\end{align}
In Eqs. (\ref{Is}, \ref{Ij}) we have identified the Fourier transformation in
time for $L$. This was the reason for separating the quantities $W_i$ from
the time-evolution. 
In order to quarantee the convergence of the transformation, an
infinitesimal (positive) convergence factor $\eta^+$ has been introduced.
Thus, we see that the modified detunings
defined already in equation \eqref{moddet} 
now appear in the expression for the current. 
Here we may identify the spatial integration as a Fourier transformation
with respect to zero momentum. 
Furthermore, in a homogeneous
system we may drop the variable $\rry$ since the current is not a function
of position. Therefore we obtain
\begin{align}
 I^{S}_{13}(t)&= -2
\im\left[\Omega_{13}^*\Omega_{13} L_{1331}(\p=0 ,\deti_{13}+i\eta^+)\right], \nonumber\\
 I^{J}_{13}(t)&= -2
 \im\left[e^{-i(\deti_{13} + \deti_{24})t} \Omega_{13}^*\Omega_{24}^* L_{1234}(\p=0 , -\deti_{24}+i\eta^+)\right]. \label{Ijpnolla}
\end{align}
We note here that the definitions of single-particle and Josephson
currents agree with their definitions in the conventional case of two
superconductors when $\deti_{13}=\deti_{24}$. 

At this point in our calculation we have isolated the effect of the
perturbation on the initial system in the linear response
approximation. Hereafter we only need to work with
linear response functions which depend only on
the properties of the unperturbed system.\\

{\centering\subsection*{Kadanoff-Baym formalism}}
 \label{sec: kb}
 Our goal here is the calculation of the retarded linear response
 function $L$. To this aim we will resort to the Kadanoff-Baym
 formalism \cite{baym_kad_1,baym} to which we refer also as the
 self-consistent method. The interest of this method for our purposes
 lies in the fact that it allows the inclusion of the
 cross-interaction effects in the transport properties, and that, on
 more general grounds, the linear response function $L$ will obey the
 same conservation laws that are obeyed by the single-particle
 Green's function $\mathcal{G}$.
 
 In the Kadanoff-Baym formalism, when the general case of a time and
 position dependent external perturbation is considered, $L$ is
 obtained from $\mathcal{G}$ by functional differentiation with
 respect to the external field

 \begin{equation} 
    L_{ijkl}(\mathbf{r}_i \tau,\mathbf{r}_j \tau,\mathbf{r}_{k} \tau^\prime,\mathbf{r}_{l} \tau^\prime)=
    \unolla{\frac{\delta\mathcal{G}_{ik}(\rrs{i} \tau,\rrs{k}\tau^\prime)}
      {\delta \Omega_{lj}(\rrs{l}\tau^\prime,\rrs{j}\tau)} }.
  \label{defL}
 \end{equation}
 Here we are working in complex time Matsubara formalism, which is a 
 standard technique for dealing with finite temperature Green's functions.
 The real-time retarded linear response function can be obtained from
 $L$ in the Matsubara formalism by analytical continuation.

For the sake of brevity, we truncate the variables such as $\rr_i$ and the spin-index
in the following two equations to just the index $i$. Furthermore, we denote the variables
of integration with a bar over the variable.
The essential starting point is the equation of motion for the single particle
Green's function
\be
\int \mathcal{G}_0^{-1}(1,\bar{1})\mathcal{G}(\bar{1},1^\prime)
=\delta(1-1^\prime)+\int \Omega(1,\bar{1})\mathcal{G}(\bar{1},1^\prime)+\int \Sigma(1,\bar{1})\mathcal{G}(\bar{1},1^\prime),
\ee
Here $\mathcal{G}_0^{-1}$ is the inverse free propagator, which accounts for the kinetic
part of the equation of motion, 
and the self-energy $\Sigma$ contains all the interaction effects in the system.
On the level of principle one should solve this equation for
an arbitrary perturbation $\Omega$ to obtain $\mathcal{G}$
and then further calculate $L$. This is in practice unfeasible, and in the Kadanoff-Baym
method one circumvents this problem by taking the variational derivative of the equation 
of motion at $\Omega=0$. This leads to an implicit equation for $L$ 
\begin{align}
L(12,1^\prime 2^\prime) = &
\int \mathcal{G} (1,\bar{3} )_{\Omega=0}  \mathcal{G}(\bar{4},1^\prime)_{\Omega=0} \unolla{\varu{\Omega(\bar{3},\bar{4})}{2^\prime}{2}}  
\nonumber\\
+& \int \mathcal{G} (1,\bar{3} )_{\Omega=0}  \mathcal{G}(\bar{4},1^\prime)_{\Omega=0}   
\unolla{\varG{\Sigma(\bar{3},\bar{4})}{\bar{5}}{\bar{6}}} L(\bar{5} 2,\bar{6} 2^\prime) , \label{Leqq}
\end{align}
which is easier to solve. In fact, in a homogeneous system we may utilize Fourier
transformations leading to a matrix equation for $L$ in momentum space.

In the following calculation we consider the case where $U_{13}\neq0$;
there is no conceptual difference when
also other cross interactions are considered different from zero. If
the BCS approximation is considered for the unperturbed system, the
explicit form of the linear response function becomes
\begin{align}
L_{1331}(\mathbf{p},\omega)&=\frac{\Pi_{\mathcal{G}}(\mathbf{p},\omega)}{1+U_{13}\Pi_{\mathcal{G}}(\mathbf{p},\omega)},\nonumber\\
L_{1234}(\mathbf{p},\omega)&=-\frac{\Pi_{\mathcal{F}}(\mathbf{p},\omega)}{1+U_{13}\Pi_{\mathcal{G}}(\mathbf{p},\omega)} \label{joseg1}
\end{align}
for the current between states $\left|1\right\rangle$ and $\left|3\right\rangle$ and
\begin{align}
L_{2442}(\mathbf{p},\omega)&=\Pi_{\mathcal{G}}(\mathbf{p},\omega)-\frac{U_{13}\Pi_{\mathcal{F}}(\mathbf{p},\omega)^2}{1+U_{13}\Pi_{\mathcal{G}}^\prime(\mathbf{p},\omega)},\nonumber\\
L_{2143}(\mathbf{p},\omega)&=-\frac{\Pi_{\mathcal{F}}(\mathbf{p},\omega)}{1+U_{13}\Pi_{\mathcal{G}}^\prime(\mathbf{p},\omega)} \label{joseg2}
\end{align}
for the current between states $\left|2\right\rangle$ and $\left|4\right\rangle$.

The terms $\Pi$ in the previous equations are
\begin{eqnarray}
  \label{eq:pi_anom}
  \Pi_{\mathcal{G}}(\mathbf{p},\omega)=\frac{1}{\beta V}
           \sum_{\mathbf{q}\,\chi}
           \mathcal{G}_{11}(\mathbf{q},\chi) \mathcal{G}_{33}(\mathbf{q}-\mathbf{p},\chi-\omega),\\
  \Pi_{\mathcal{G}}^\prime(\mathbf{p},\omega)=\frac{1}{\beta V}
           \sum_{\mathbf{q}\,\chi}
           \mathcal{G}_{33}(\mathbf{q},\chi) \mathcal{G}_{11}(\mathbf{q}-\mathbf{p},\chi-\omega),\\
   \Pi_{\mathcal{F}}(\mathbf{p},\omega)=\frac{1}{\beta V}
           \sum_{\mathbf{q}\,\chi}
           \mathcal{F}_{12}(\mathbf{q},\chi) \mathcal{F}^*_{34}(\mathbf{q}-\mathbf{p},\chi-\omega).
\end{eqnarray}
Here $\mathcal{F}$ is the so called anomalous Green's function,
which expresses the pair correlation of the superfluid.
In the main article we study a case where the two superfluids are identical, in which
case we have $\Pi_\mathcal{G}=\Pi_\mathcal{G}^\prime$.

The expression for the linear response function leads to the following
results for the single-particle and Josephson currents
\begin{align}
I^{S}_{13}( t) =& -2 |\Omega_{13}|^2 
\times\im \Bigg[ \frac{\Pi_{\mathcal{G}}(\p=0 ,\deti_{13}+i\eta^+)}{1+U_{13}\Pi_{\mathcal{G}}(\p=0 ,\deti_{13}+i\eta^+)}\Bigg], \label{eq:KBsingle1}\\
I^{S}_{24}( t) =& -2 |\Omega_{24}|^2 
\times\im\Bigg[  \Pi_{\mathcal{G}}(\p=0 ,\deti_{24}+i\eta^+)
-\frac{U_{13}\Pi_{\mathcal{F}}(\p=0 ,\deti_{24}+i\eta^+)^2}{1+U_{13}\Pi_{\mathcal{G}}^\prime(\p=0 ,\deti_{24}+i\eta^+)}\Bigg] . \label{eq:KBsingle2}
\end{align}
\begin{align} 
I^{J}_{13}( t) =&2 \left|\Omega_{13}\Omega_{24}
\frac{\Pi_{\mathcal{F}}(\p=0 , -\deti_{24}+i\eta^+)}{1+U_{13}\Pi_{\mathcal{G}}(\p=0 ,-\deti_{24}+i\eta^+)}\right|
\times
\sin[(\deti_{13} + \deti_{24})t+\varphi(\deti_{24})], \label{eq:KBjose1}\\
I^{J}_{24}( t) =&2 \left|\Omega_{13}\Omega_{24}
\frac{\Pi_{\mathcal{F}}(\p=0 , -\deti_{13}+i\eta^+)}{1+U_{13}\Pi_{\mathcal{G}}^\prime(\p=0 ,-\deti_{13}+i\eta^+)}\right|
\times
\sin[(\deti_{13} + \deti_{24})t+\varphi(\deti_{13})]. \label{eq:KBjose2}
\end{align}
The phase factors $\varphi(\deti_{13})$ and $\varphi(\deti_{24})$ which appear
in equations \eqref{eq:KBjose1} and \eqref{eq:KBjose2} are the
initial complex phase of the retarded linear response function $L$.
In the article we have left out the minus sign from the detunings in the terms $\Pi$
since in the case of identical superfluids, which we consider in the article,  
the critical current becomes a symmetric function of the detuning.\\

{\centering\subsection*{Identical superfluids}}
\label{sec:Js_phi}

Here we discuss the results obtained in the case of identical superfluids. 
In this case the notation simplifies as 
$\tilde{\delta}_{13}=\delta_{13}+\mu_3-\mu_1=\delta_{13}$ and 
$\tilde{\delta}_{24}=\delta_{24}+\mu_4-\mu_2=\delta_{24}$
since the chemical potentials are equal.
From the Kadanoff-Baym formalism, in the case where
cross-interactions are zero, we have
\begin{align}
I^{S}_{13}(\rry t) &= -2 |\Omega_{13}|^2 \im\left[ \Pi_{\mathcal{G}}(\p=0 ,\delta_{13}+i\eta^+)\right], \nonumber\\
I^{S}_{24}(\rry t) &= -2 |\Omega_{24}|^2 \im\left[ \Pi_{\mathcal{G}}(\p=0 ,\delta_{24}+i\eta^+)\right], \label{Isid}
\end{align}
and
\begin{align} 
I^{J}_{13}(\rry t) &= 2 \left|\Omega_{13}\Omega_{24}
\Pi_{\mathcal{F}}(\p=0 , -\delta_{24}+i\eta^+)\right|\sin\left((\delta_{13} + \delta_{24})t +\varphi_1(\delta_{24})\right), \nonumber\\
I^{J}_{24}(\rry t) &= 2 \left|\Omega_{13}\Omega_{24}
\Pi_{\mathcal{F}}(\p=0 , -\delta_{13}+i\eta^+)\right|\sin\left((\delta_{13} + \delta_{24})t +\varphi_2(\delta_{13})\right). \label{Ijid}
\end{align}
 \begin{figure}
    \centering
    \epsfig{file=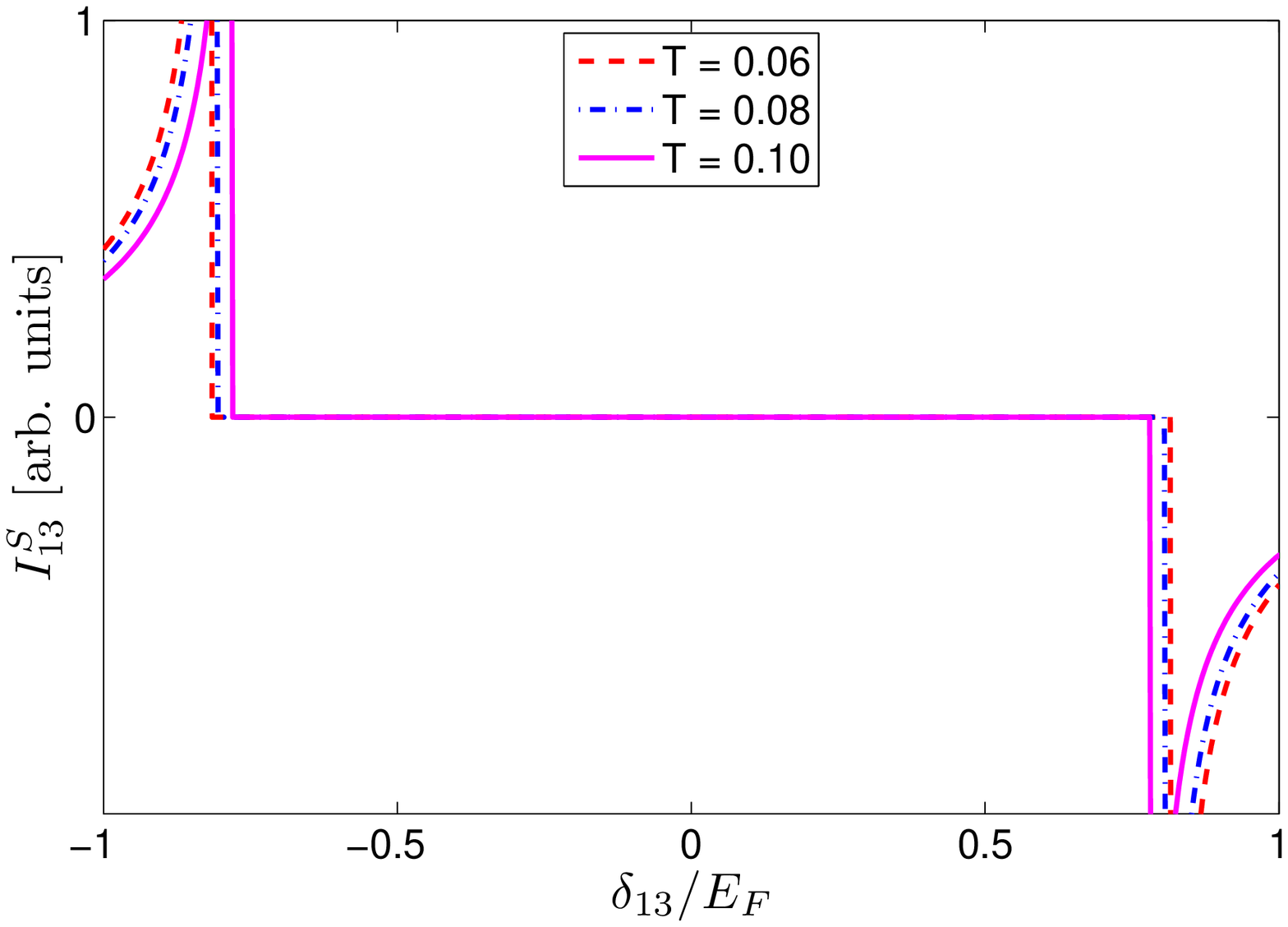,width=0.3\columnwidth}
    \epsfig{file=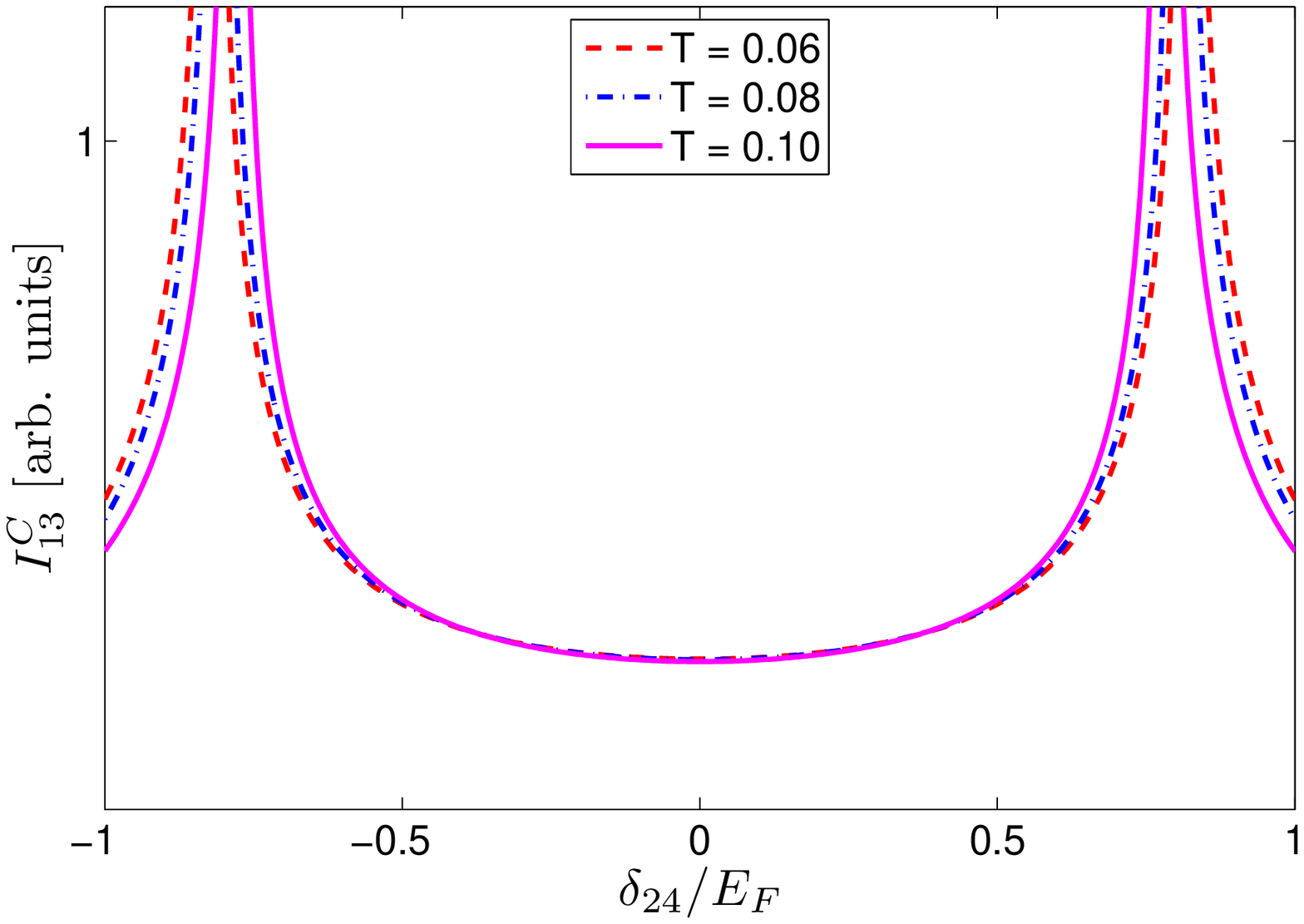,width=0.3\columnwidth}
    \caption{(left to right). The single-particle current $I_{13}^S$ and the
      critical Josephson current $I_{13}^C$ \textit{vs.} ${\delta}_{13}$ and
      ${\delta}_{24}$ for identical BCS states. $U_{12}=U_{34}=-2.0E_F/n$. 
      $T$ is in the units of Fermi temperature $T_F$ and in the temperature range
      of the figure, the gap is $\Delta\sim 0.4 E_F$.}
  \label{fig:symm}
  \end{figure}
  The single-particle current vanishes when $|\delta_{ij}|<2\Delta_{12}$ (Fig. \ref{fig:symm}).  
  In this range only the Josephson current exists, as it is intuitively clear, since the
  single-particle current requires the breaking of Cooper pairs.
  At $|\delta_{ij}|=2\Delta_{12}$ the detuning equals the minimum of
  the transition energy $E_{12}(\q)+E_{34}(\q)$, where
  $E_{\alpha\beta}(\q)=\sqrt{\vas|q|^2/2m-\mu_\alpha\oik^2+|\Delta_{\alpha\beta}|^2}$.
  This occurs at the point $\frac{q^2}{2m}=\mu$
  \textit{i.e.} at the Fermi level, where the BCS density of states is divergent,
  leading to a divergence in the dependence of the currents on the
  detuning. In the case of the critical Josephson current, this
  divergence is commonly known as the Riedel peak.

  To illustrate the spin-asymmetry further, consider the results of Fig. \ref{fig:symm}.
  Notice that we plot only $I^C_{13}$ since $I^C_{24}$ is the same function, the difference being only
  that $I^C_{13}$ depends on $\delta_{24}$ whereas $I^C_{24}$ depends on $\delta_{13}$.
  For instance taking $\delta_{13}/E_F=0.4$ and $\delta_{24}/E_F=0.5$ at $T/T_F=0.08$ then leads to 
  $I^C_{13}(\delta_{24})/I^C_{24}(\delta_{13})=1.14$, in other words, a 14 \% difference in the amplitude of
  the Josephson current for each spin component.\\

\begin{figure}
    \centering
    \epsfig{file=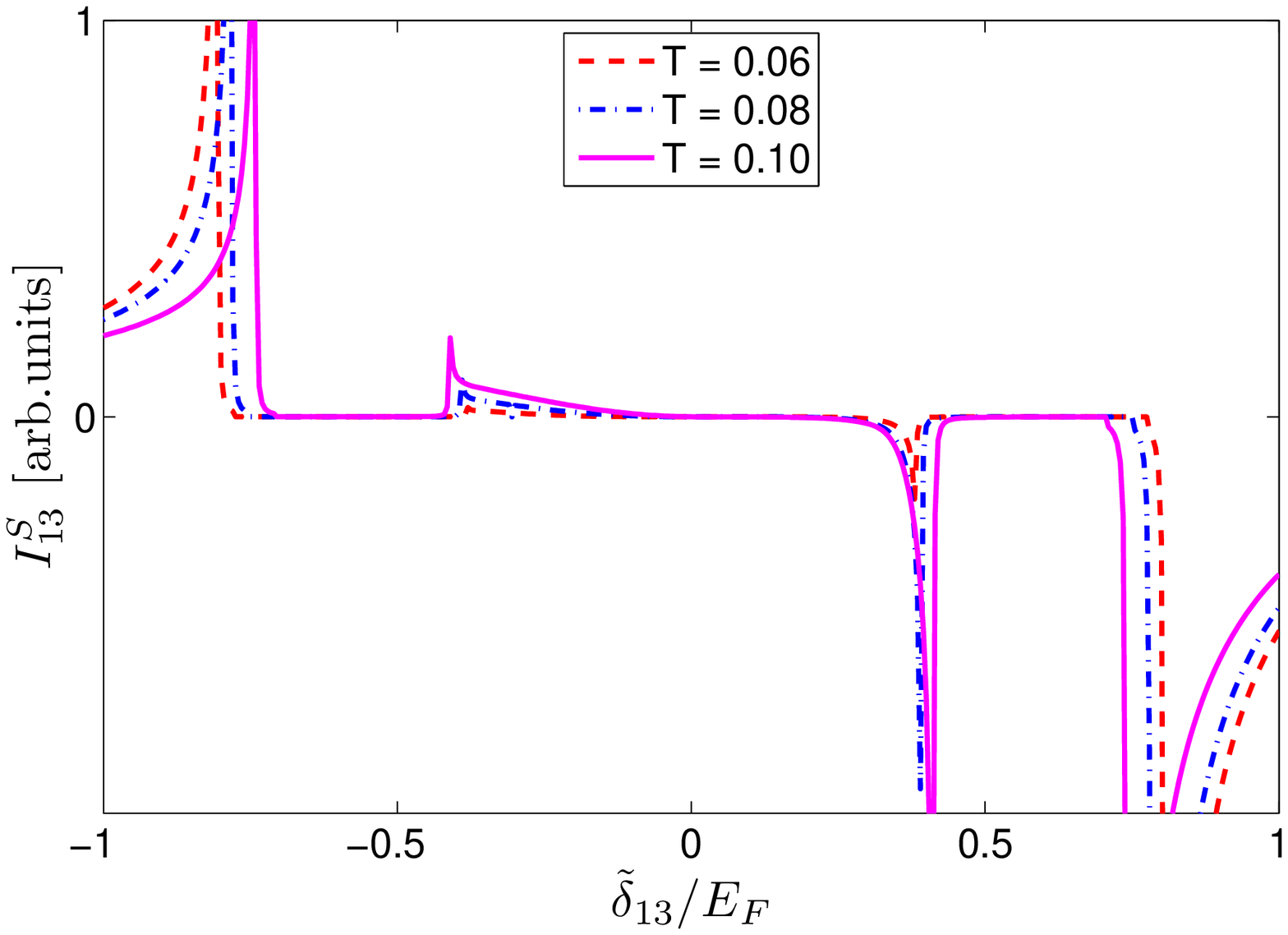,width=0.3\columnwidth}
    \epsfig{file=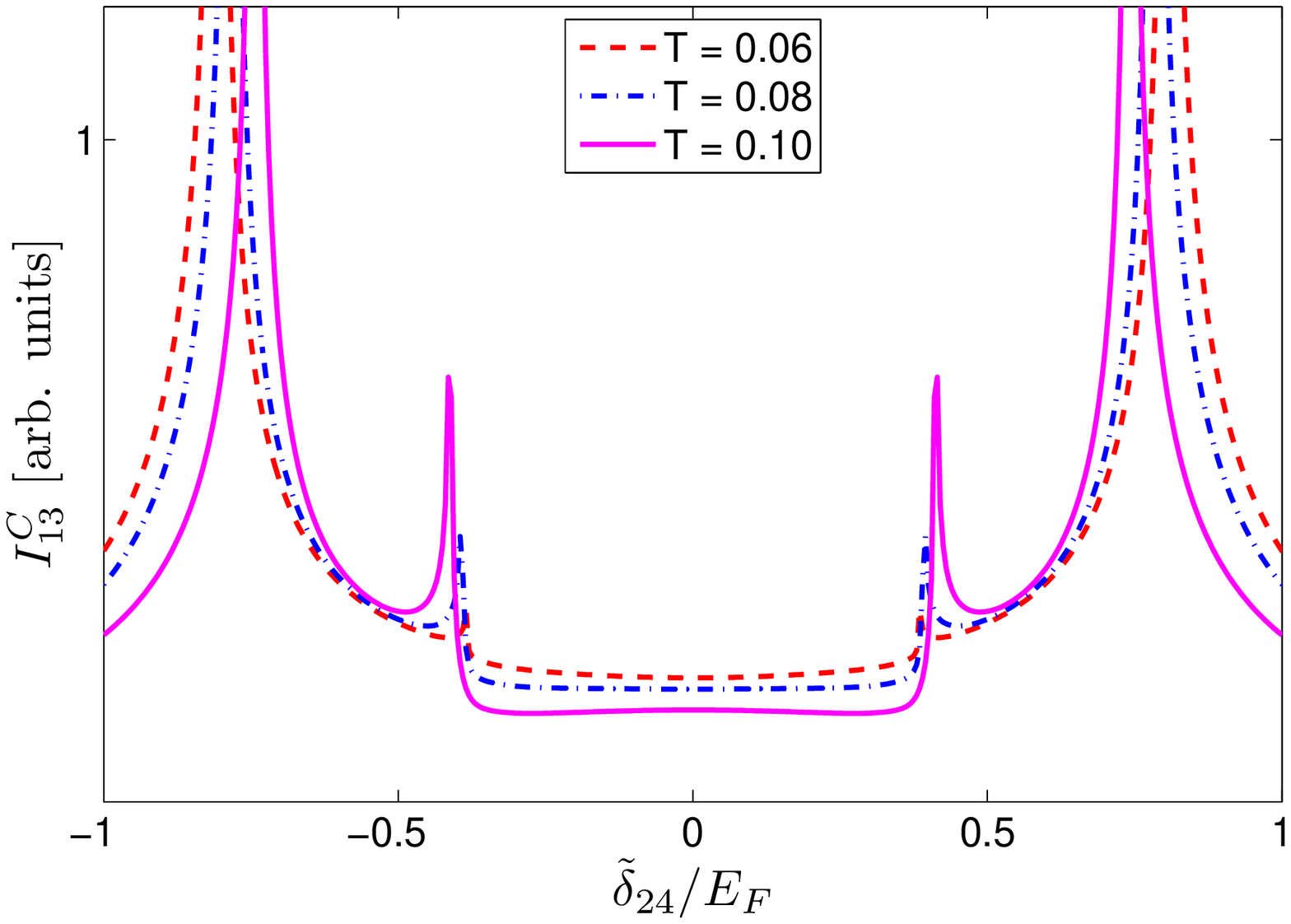,width=0.3\columnwidth}
    \caption{(left to right). The single-particle current
      $I_{13}^S$ and the critical Josephson current $I_{13}^C$  \textit{vs.}
      $\tilde{\delta}_{13}$ and $\tilde{\delta}_{24}$ 
for unidentical BCS states. $U_{12}=-1.0E_F/n$, $U_{34}=-5.0E_F/n$, $T$ is in the units of Fermi temperature $T_F$.}
  \label{fig:asymm}
\end{figure}

\begin{figure}
    \centering
    \epsfig{file=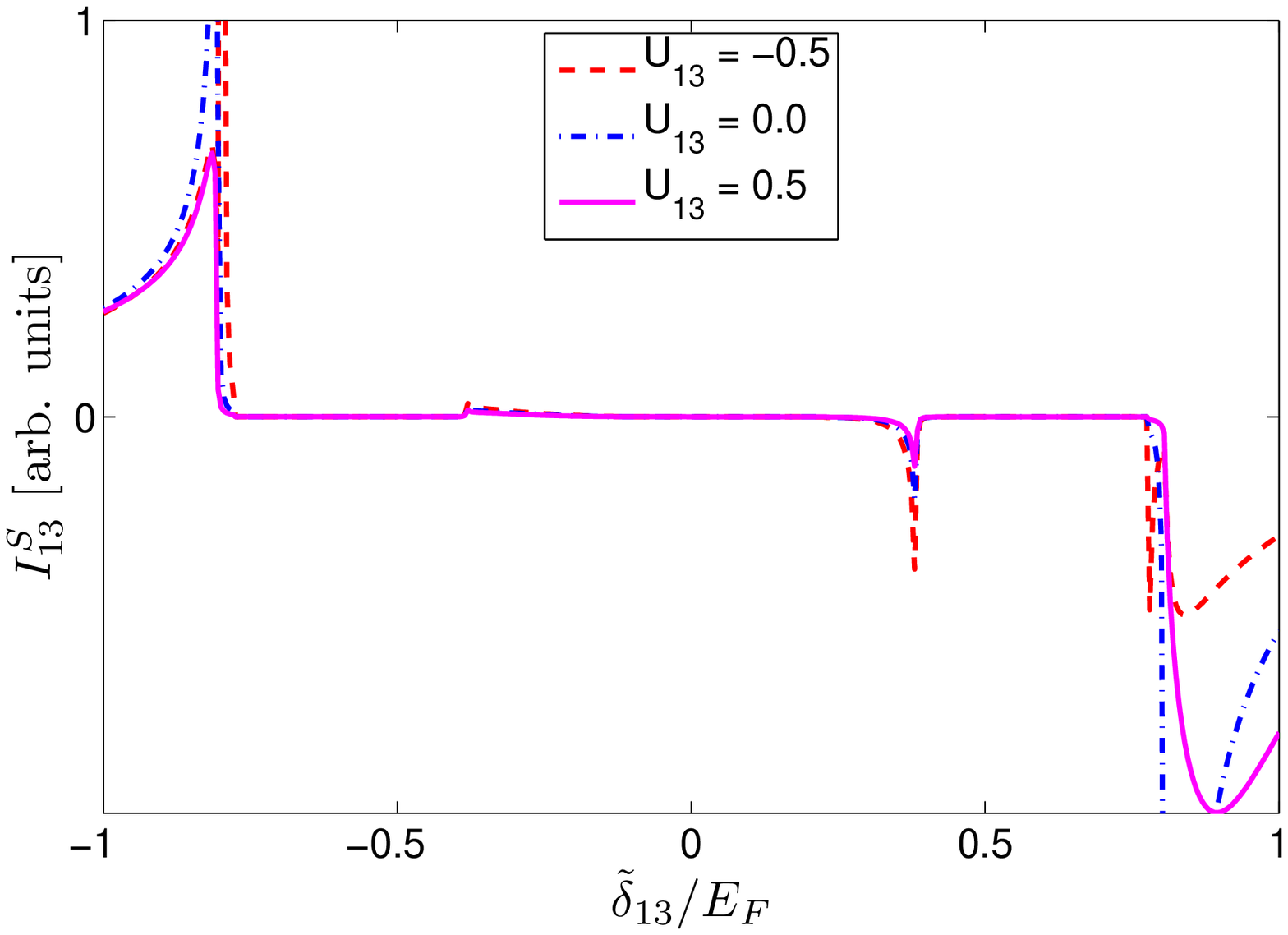,width=0.3\columnwidth}
    \epsfig{file=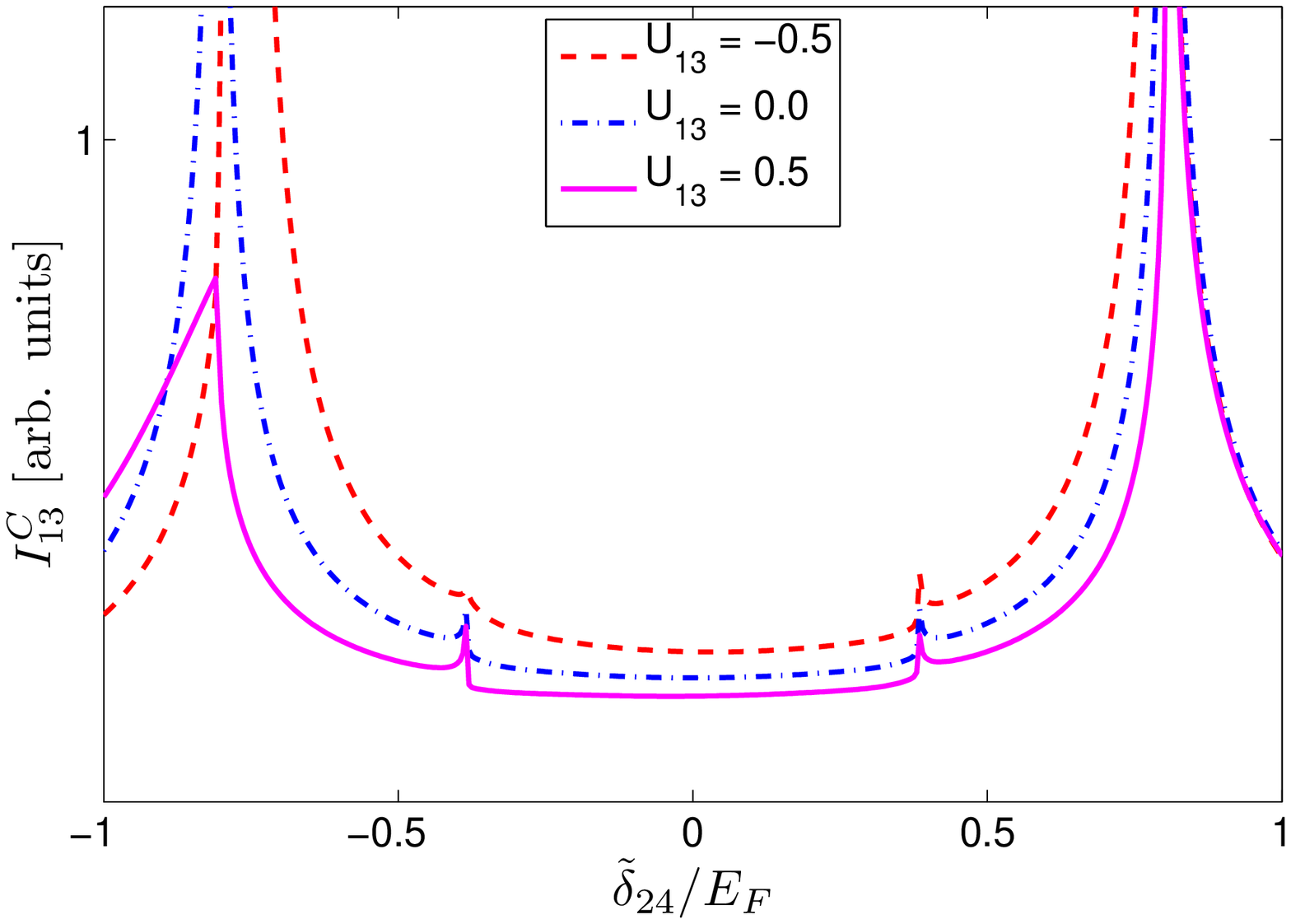,width=0.3\columnwidth}
    \caption{Single-particle current
      $I_{13}^S$, and critical Josephson current $I_{13}^C$ \textit{vs.}
      $\tilde{\delta}_{13}$ and  $\tilde{\delta}_{24}$ 
      for unidentical BCS states. $U_{12}=-1.0E_F/n$, $U_{34}=-5.0E_F/n$, $T$ is in the units of Fermi temperature $T_F$.}
  \label{fig:cross}
\end{figure}

{\centering\subsection*{Finite temperature and cross interactions}}
\label{sec:T_cross}
It is instructive and of certain interest, from the experimental point
of view, to discuss how the results presented in the article are affected
by temperature and different interaction strengths.  In order to
properly observe finite temperature effects, we consider the
slightly more complicated case of two different BCS states. 
Experimentally, the condensates 1-2 
and 3-4 may have different interaction strenghts. We choose
$U_{12}=-1.0E_F/n$ and $U_{34}=-5.0E_F/n$. This leads to gaps $\Delta_{12} \sim 0.2 E_F$ 
and $\Delta_{34} \sim 0.5 E_F$, 
which corresponds roughly to systems that have been realised experimentally.
Also the choice of temperature we use, $T=0.06\ldots 0.1 T_F$, is motivated by the lowest
temperatures reached in the experiments. The results are plotted in Figs. \ref{fig:asymm} and \ref{fig:cross}.
First of all we notice that all the features previously found still dominate the
result, even though there are some clear changes.  

We see that for finite $T$ and $U_{12}\neq U_{34}$ there can be a
finite single particle current even at small detunings.
This current arises due to thermal excitations in the BCS states.
Intuitively, this contribution is generated by Cooper pairs that have
been already broken because of thermal energy. To be more precise, it
is the fact that one of the BCS states has more thermal excitations
for a given energy than the other that leads to the emergence of the
thermal single particle current.  We did indeed have thermal
excitations in the results of the previous section as well,
but their net contribution summed up to zero. The new peaks in
the results (see Figures \ref{fig:asymm} and \ref{fig:cross}) are associated with the the
contribution of thermal excitations diverging at a particular detuning.
The explanation for these divergences is similar as for the divergences
encountered in the previous section.  Here the energy of the
tunneling process is $E_{12}(\q)-E_{34}(\q)$ 
(again with $E_{\alpha\beta}(\q)
=\sqrt{\vas\frac{1}{2m}|q|^2-\mu_\alpha\oik^2+|\Delta_{\alpha\beta}|^2}$) and
the resonance occurs when the detuning coincides with the extremum of
this energy.  The density of states argument of the previous section holds
then here as well.  Moreover, the thermal single particle current
vanishes above this maximum because at larger detunings there are no
thermal transitions that would conserve both energy and momentum.\\

\begin{figure}
    \centering
    \epsfig{file=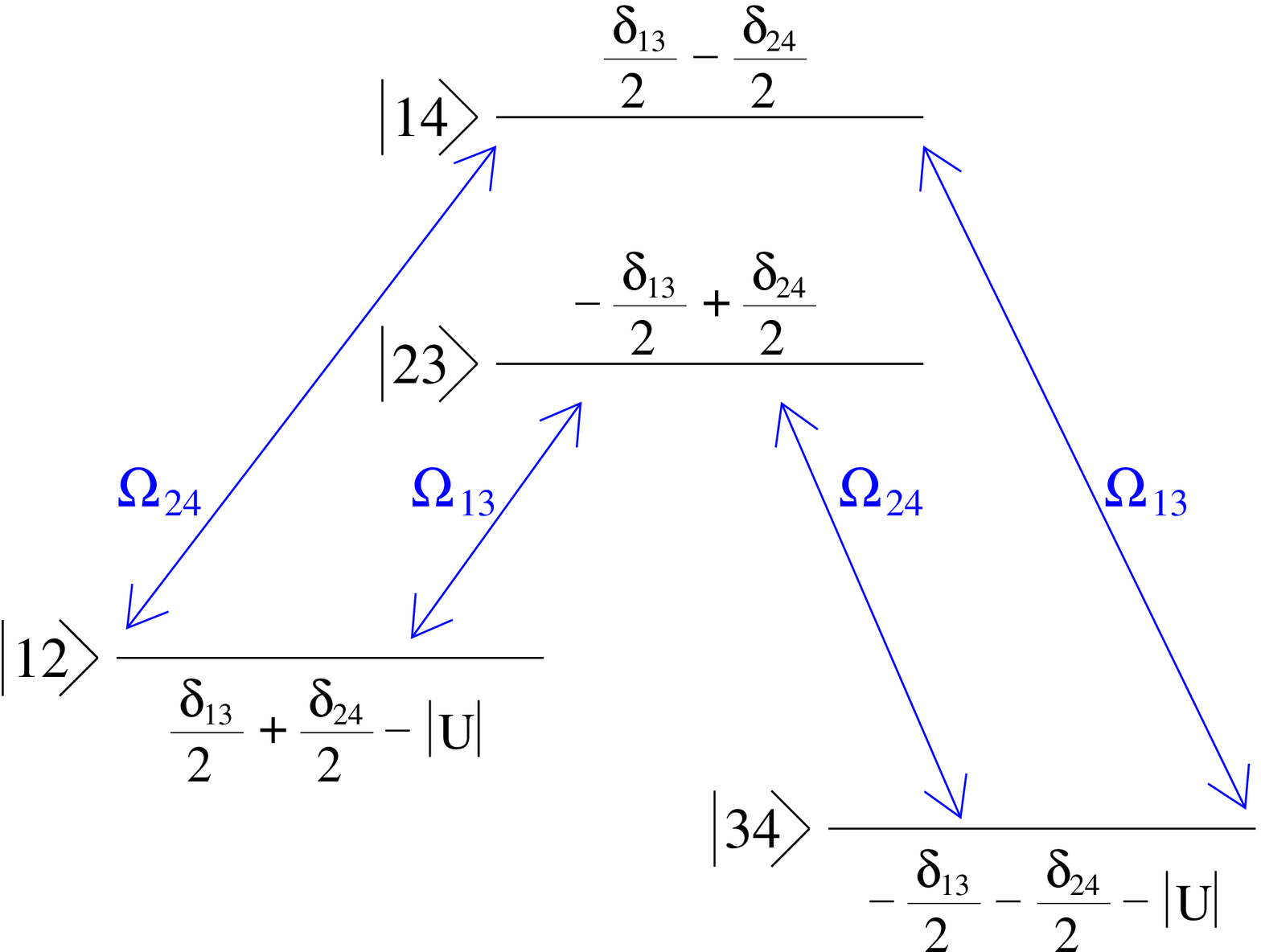,width=0.6\columnwidth}
    \caption{Energy levels for the four-state model.}
  \label{fig:level_scheme}
  \end{figure}

\newpage
{\centering\section*{The perturbative calculation for the four-state system}}
\label{sec:pert}
{\centering\subsection*{The Hamiltonian}}
For the perturbative calculation of the four-state system of Figure \ref{fig:level_scheme}, 
we consider the following Hamiltonian
\begin{equation}
  \label{eq:Hamiltonian}
   H=H_{0} + H_{RF},
\end{equation}
where $H_{0}$ contains the energies of the states and $H_{RF}$ is the RF coupling. They have the form
\begin{align}
  \label{eq:Hamiltonian2}
  H_{0}&= 
   -U \left( n_{ 1} n_{ 2} +  n_{ 3} n_{ 4} \right),\nonumber\\
  H_{RF}&=\Omega_{13}  c_{ 1}^\dagger c_{ 3} +\textit{h.c.}
       +\Omega_{24}  c_{ 2}^\dagger c_{ 4} + \textit{h.c.} 
        + \frac{\delta_{13}}{2}\left( n_{ 1}-n_{ 3} \right) 
        + \frac{\delta_{24}}{2}\left( n_{ 2}-n_{ 4} \right) ,
\end{align}
where $U$ is the energy difference between the paired and unpaired states. We have
kept the notation here as similar as possible to the initial description in the article.
Obviously one could reformulate this four-state system in a more compact form.

{\centering\subsection*{Perturbative calculation}}

Now let us study our model system analytically using perturbation theory.
The dynamics of the system can be described by
four states: the paired states $\ket{\phi_{\rm I}}\equiv\left|12\right\rangle$, 
$\ket{\phi_{\rm II}}\equiv\left|34\right\rangle$ -- corresponding to the left and right
well respectively in the double well description -- and the states
$\left|14\right\rangle$, $\left|23\right\rangle$ (see Figure
\ref{fig:level_scheme}).

In order to find the lowest order contribution of the RF-couplings we 
have to employ
the time-dependent perturbation theory to the second order in the RF-couplings.
As an initial state we consider an arbitrary superposition of the paired states
\begin{equation}
  \label{eq:state}
  \left|\phi_0\right\rangle=
    \alpha_0\left|\phi_{\rm I}\right\rangle+\beta_0\left|\phi_{\rm II}\right\rangle.
\end{equation}
We point out already that in the  $\alpha_0=0$ or $\beta_0=0$
case no Josephson current will appear. 
Calculating the occupation number  $N_1$ of the
hyperfine state in the state $\left|\phi(t)\right\rangle =
\exp\left[-iHt\right]\left|\phi_0\right\rangle$ gives
\begin{eqnarray}
  \left\langle\phi(t)|N_1|\phi(t)\right\rangle&=& |\alpha_0|^2 \left\langle\phi_{\rm I}(t)|N_1|\phi_{\rm I}(t)\right\rangle+ |\beta_0|^2 \left\langle\phi_{\rm II}(t)|N_1|\phi_{\rm II}(t)\right\rangle 
       \nonumber\\
  &&  + \alpha_0 \beta_0^*\left\langle\phi_{\rm II}(t)|N_1|\phi_{\rm I}(t)\right\rangle + \alpha_0^* \beta_0\left\langle\phi_{\rm I}(t)|N_1|\phi_{\rm II}(t)\right\rangle.
   \label{eq:nump_1}
\end{eqnarray}
The perturbative expansion for $\left|\phi_{\rm I}(t)\right\rangle$ and
 $\left|\phi_{\rm II}(t)\right\rangle$ reads
\begin{eqnarray}
   \left|\phi_{\rm I}(t)\right\rangle&=&\left|12\right\rangle + 
                 \gamma_{12}^{(1)}(t) \left|14\right\rangle + 
                 \delta_{12}^{(1)}(t) \left|23\right\rangle + 
                 [ \alpha_{14\,12}^{(2)}(t)+ \alpha_{23\,12}^{(2)}(t)]  \left|12\right\rangle + 
                 [ \beta_{14\,12}^{(2)}(t)+ \beta_{23\,12}^{(2)}(t)]  \left|34\right\rangle \nonumber \\
   \left|\phi_{\rm II}(t)\right\rangle&=&\left|34\right\rangle + 
                 \gamma_{34}^{(1)}(t) \left|14\right\rangle + 
                 \delta_{34}^{(1)}(t) \left|23\right\rangle + 
                 [ \alpha_{14\,34}^{(2)}(t)+ \alpha_{23\,34}^{(2)}(t)]\left|12\right\rangle + 
                 [ \beta_{14\,34}^{(2)}(t)+ \beta_{23\,34}^{(2)}(t)]\left|34\right\rangle
   \label{eq:pert_st}
\end{eqnarray}
in equation \eqref{eq:pert_st} the superscript represents the
perturbative order, the subscript the path corresponding to the
process considered and the Greek letter its starting state (with the
convention $\alpha\leftrightarrow\ket{12}$,
$\beta\leftrightarrow\ket{34}$,$\gamma\leftrightarrow\ket{14}$,$\delta\leftrightarrow\ket{23}$).
For instance $\beta_{14\,12}^{(2)}$ corresponds to the second-order
process $\ket{34}\to\ket{14}\to\ket{12}$. The coefficients that we need
for calculating $\left\langle\dot{N}_1\right\rangle$
\begin{align}
  \gamma_{12}^{(1)}(t)&=2 i \frac{\Omega_{24}}{\omega_{14\,12}}
                   \sin(\omega_{14\,12}t/2)\exp\left(i\omega_{14\,12}t/2\right) \nonumber\\
   \label{eq:pert_coeff}
   \alpha_{14\,34}^{(2)}(t)&=-4\Omega_{13}\Omega_{24}
                      \left[
                            \frac{\exp[i(\omega_{14\,34}-\omega_{14\,12})t/2]\sin\left[(\omega_{14\,34}-\omega_{14\,12})t/2\right]}
                                 {2i\omega_{14\,34}(\omega_{14\,34}-\omega_{14\,12})} -
                            \frac{\exp(-i\omega_{14\,12}t/2)\sin(\omega_{14\,12}t/2)}{2i\omega_{14\,12}\omega_{14\,34}}
                      \right]
\end{align}
are obtained within time-dependent perturbation theory, considering
$H_0$ as the initial state and switching on $H_{RF}$ at $t=0$. In the
equation above we have defined
\begin{eqnarray}
  &&\omega_{14\,12}= U-\delta_{24} \label{eq:omegas1}\\
  &&\omega_{23\,12}= U-\delta_{13} \label{eq:omegas2}\\
  &&\omega_{14\,34}= U+\delta_{13} \label{eq:omegas3}\\
  &&\omega_{23\,34}= U+\delta_{24}.\label{eq:omegas4}
\end{eqnarray}

Since we are interested in the time evolution of the expectation value
of the particle number in state $\left| 1 \right\rangle$ up to second
order in $\Omega$, the only relevant second-order coefficients are
$\alpha_{14\,34}^{(2)}(t)$ and $\alpha_{23\,34}^{(2)}(t)$. This is due
to the fact that the states corresponding to other 2$^{nd}$ order
coefficients do not contribute to $\left\langle N_1\right\rangle$
through the scalar product with $\left|12\right\rangle$ (0$^{th}$
order term).  The calculation of the terms
$\left\langle\phi_{\rm I}(t)|N_1|\phi_{\rm I}(t)\right\rangle$ and
$\left\langle\phi_{\rm II}(t)|N_1|\phi_{\rm II}(t)\right\rangle$ give the
quasiparticle current ($I^s_{ij}$ in equation (1) of the main article) 
when the proper normalization is taken into account.

In the term $\alpha_0 \beta_0^*\left\langle\phi_{\rm II}(t)|N_1|\phi_{\rm I}(t)\right\rangle 
+ \alpha_0^* \beta_0\left\langle\phi_{\rm I}(t)|N_1|\phi_{\rm II}(t)\right\rangle$
we have contributions both from the scalar product between 1$^{st}$ order terms 
($\left\langle14\left| \dots \right|14\right\rangle$)
\begin{multline}
  \label{eq:1st_order}
    \alpha_0^*\beta_0 \gamma_a^{(1)*}(t)\gamma_b^{(1)}(t)+h.c.= \\
       2\frac{\Omega_{13}\Omega_{24}|\alpha_0 \beta_0|}{\omega_{14\,12}\omega_{14\,34}}
        \left[
              \cos\left[(\omega_{14\,34}-\omega_{14\,12})t+\varphi\right]+\cos(\varphi)-
              \cos\left(\omega_{14\,12}t-\varphi\right)-\cos\left(\omega_{14\,34}t+\varphi\right)  
        \right]
\end{multline}
and from the scalar product between the 0$^{th}$-2$^{nd}$ order terms ($\left\langle12\left| \dots \right|12\right\rangle$)
\begin{multline}
  \label{eq:2nd_order}
  \alpha_0^*\beta_0\left[ \alpha_{14\,34}^{(2)}(t)+ \alpha_{23\,34}^{(2)}(t)\right]+h.c. =\\ 2 \Omega_{13}\Omega_{24}|\alpha_0 \beta_0|
                             \left[ 
                                  \frac{\cos\left[(\omega_{14\,34}-\omega_{14\,12})t+\varphi\right] -\cos(\varphi)}{\omega_{14\,34}(\omega_{14\,34}-\omega_{14\,12})} +
                                  \frac{\cos(\omega_{14\,12}t-\varphi)-\cos(\varphi)}{\omega_{14\,12}\omega_{14\,34}} +\right.\\
                                  \left.\frac{\cos\left[(\omega_{23\,34}-\omega_{23\,12})t+\varphi\right] -\cos(\varphi)}{\omega_{23\,34}(\omega_{23\,34}-\omega_{23\,12})} +
                                  \frac{\cos(\omega_{23\,12}t-\varphi)-\cos(\varphi)}{\omega_{23\,12}\omega_{23\,34}}
                             \right]. 
\end{multline}
Here $\varphi$ stands for the initial phase difference between $\alpha_0$ and $\beta_0$
\textit{i.e.} $\alpha_0^*\beta_0=|\alpha_0\beta_0|\exp(i\varphi)$.
In Equations (\ref{eq:1st_order}) and (\ref{eq:2nd_order}), the components
oscillating with frequency
$\left(\omega_{14\,34}-\omega_{14\,12}\right)$ correspond to the
contributions giving rise to the Josephson current, while components
oscillating at $\omega_{14\,34}$, $\omega_{23\,12}$ represent a
contribution to the quasi-particle current, and the terms in
$\omega_{14\,12}$ cancel out.  

If we now focus on the component of $\left\langle N_1\right\rangle$ oscillating at the
Josephson frequency and replace $\omega_{ij\,kl}$ with their values in
terms of interaction energies and detunings as given in equations 
(\ref{eq:omegas1}) and (\ref{eq:omegas2}), we obtain
\begin{equation}
  \label{eq:NJoseph}
   \left\langle N_1\right\rangle_J=2\Omega_{13}\Omega_{24}
                           \left[
                                 \frac{1}{(U+\delta_{13})(\delta_{13}+\delta_{24})}+
                                 \frac{1}{(U+\delta_{24})(\delta_{13}+\delta_{24})}+
                                 \frac{1}{(U-\delta_{24})(U+\delta_{13})}
                           \right]\cos[(\delta_{13}+\delta_{24})t+\varphi].
\end{equation}
Note that we would also have a contribution to the Josephson current
coming from the scalar product between the terms containing $
\alpha_{14\,12}^{(2)}(t)+ \alpha_{23\,12}^{(2)}(t)$ and $
\alpha_{14\,34}^{(2)}(t)+ \alpha_{23\,34}^{(2)}(t)$, associated with
two different paths, via two different internal states, for the
tunneling of Cooper pairs, but this is only a fourth order process.
Such processes are present even if one of the paired states would
initially be empty. 

Differentiating Equation \eqref{eq:NJoseph} with respect to time, we have
\begin{equation}
  \label{eq:NJoseph2_pre}
   I_{13}^J=2\Omega_{13}\Omega_{24}|\alpha_0\beta_0| \left[ M_{\rm pair}+ M_{\rm single} \right] 
   \sin[(\delta_{13}+\delta_{24})t+\varphi],
\end{equation}
where $M_{\rm pair}$ corresponds to the contributions to the Josephson current
coming from Equation \eqref{eq:2nd_order}, \textit{i.e.} from interference of
pair tunneling and the initial population,
\begin{equation}
  \label{eq:A}
  M_{\rm pair}= \frac{1}{(U+\delta_{13})}+\frac{1}{(U+\delta_{24})}
\end{equation}
and $M_{\rm single}$ to that coming from Equation \eqref{eq:1st_order}, \textit{i.e.} from
the interference in the single-particle tunneling to the excited state $\left|14\right\rangle$
\begin{equation}
  \label{eq:B}
  M_{\rm single}= \frac{\delta_{13}+\delta_{24}}{(U-\delta_{24})(U+\delta_{13})}.
\end{equation}
From equation (\ref{eq:NJoseph2_pre}) we obtain  the relation given in the
article for the Josephson current
\begin{equation}
  \label{eq:NJoseph3_pre}
   I_{13}^J=2\Omega_{13}\Omega_{24}|\alpha_0\beta_0| \left[ \frac{1}{(U+\delta_{24})}+
    \frac{1}{(U-\delta_{24})} \right] 
   \sin[(\delta_{13}+\delta_{24})t+\varphi] .
\end{equation} 

{\centering\subsection*{Coupling dependence of the Josephson frequency}}

\begin{figure}
    \centering
    \epsfig{file=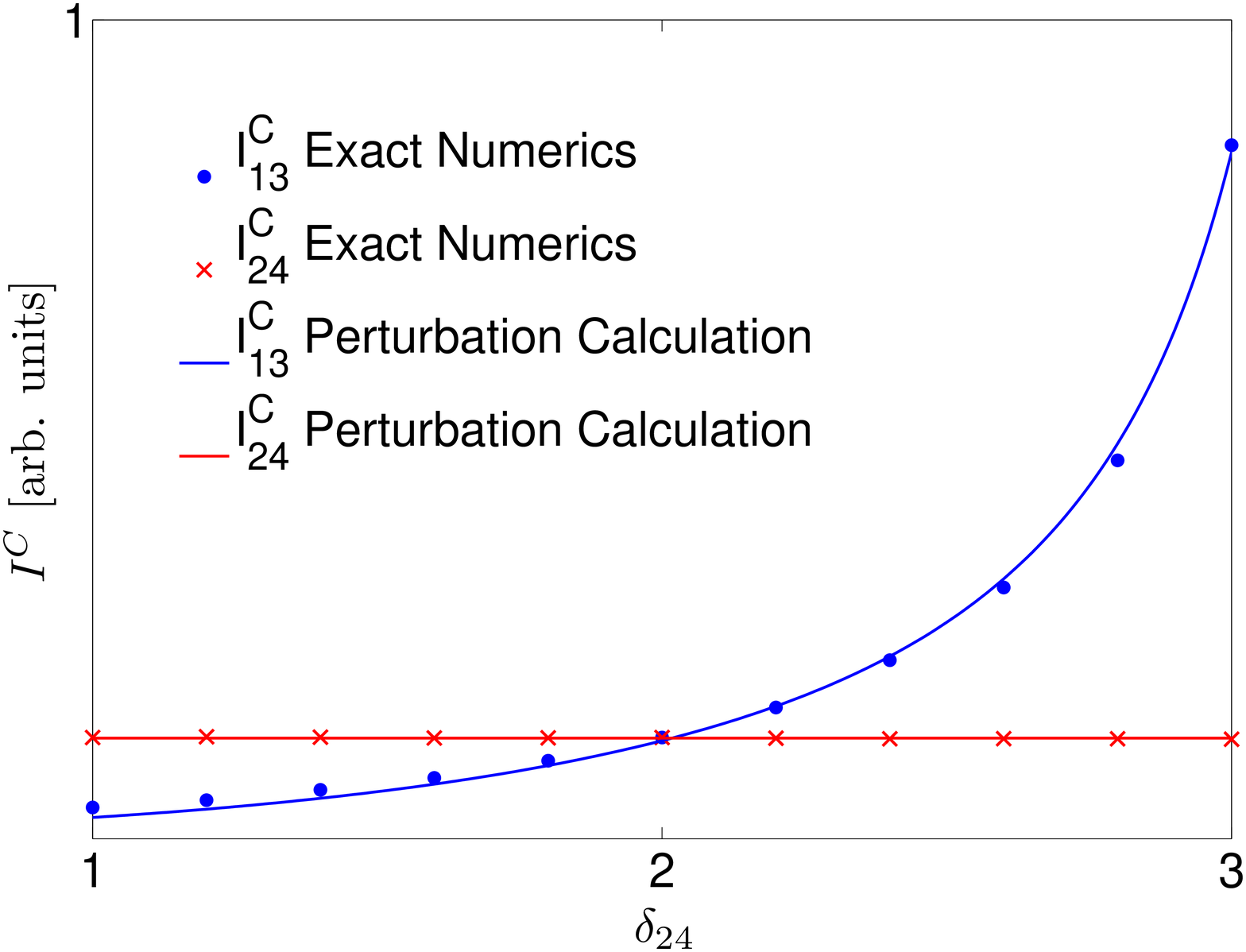,width=0.65\columnwidth}
    \caption{
The Josephson
critical currents $|I_{13}^C|$ (dots) and $|I_{24}^C|$ (crosses) as a
function of $\delta_{24}/J$, when $\delta_{13}/J=2.0$, given by the exact
numerical calculation for the four-state system (here
$U/J=5.0$ and $\Omega/J=0.1$). The solid lines show the results of the
perturbative calculation. $J$ is the unit of energy.}
  \label{fig:benchmark}
\end{figure}

\begin{figure}
    \centering
    \epsfig{file=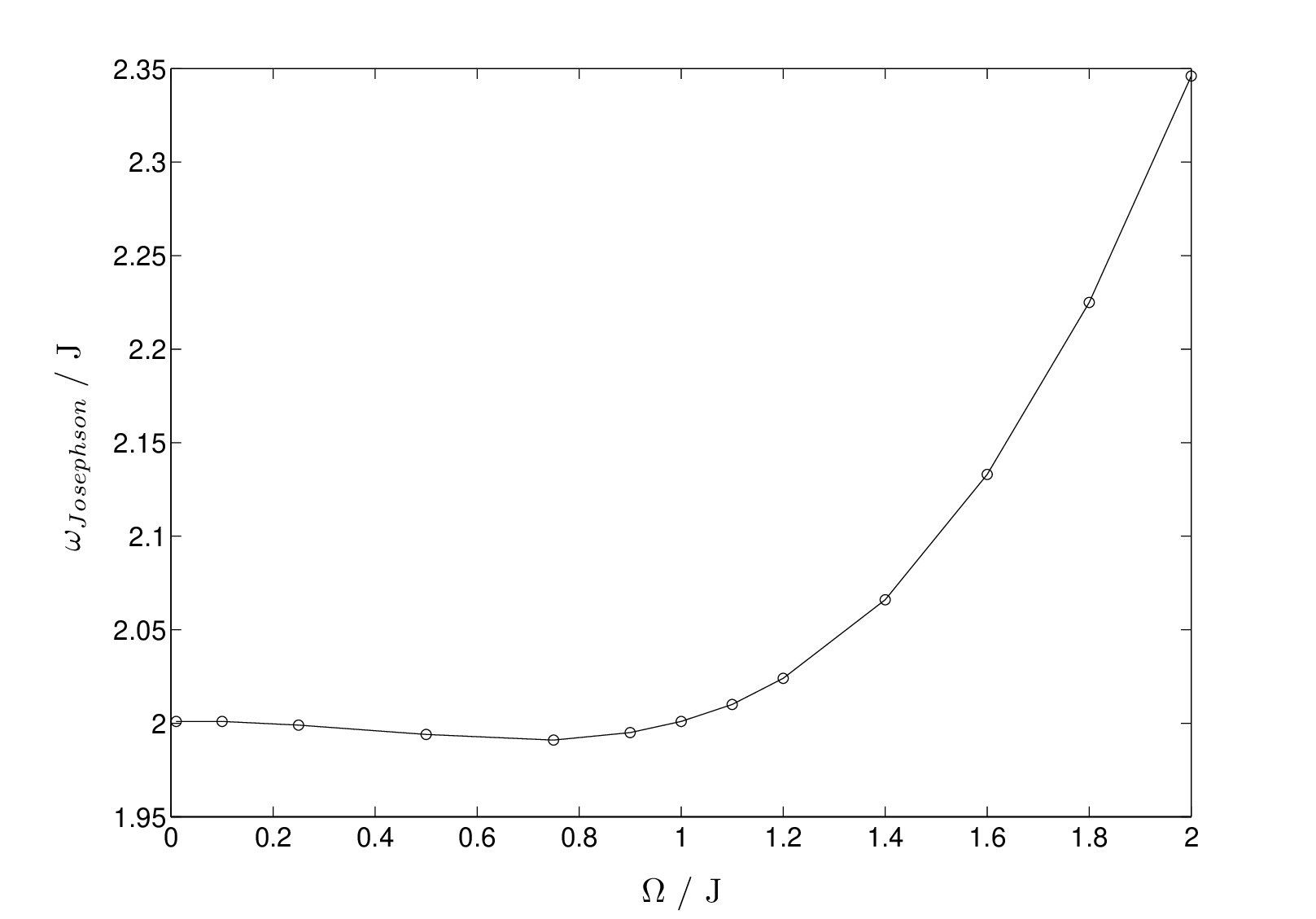,width=0.65\columnwidth}
    \caption{Position of the Josephson peak (\textit{i.e.} the Josephson
      oscillation frequency) as a function of the RF
      coupling, with $\Omega_{13}=\Omega_{24}=\Omega$, $\delta_{13}/J=\delta_{24}/J=1.0$
      and $U/J=10.0$.}
  \label{fig:JosevsOmega}
\end{figure}

The exact numerical solution of the four-state model is very much feasible,
and supplements well the perturbative analytical treatment above.
In Figure \ref{fig:benchmark} we show that for small couplings the perturbative
treatment agrees with exact numerics as one would expect.
We may then study numerically the higher order effects in the coupling strength.
The dependence of the Josephson frequency on the coupling strength is a particularly
interesting question. In Figure \ref{fig:JosevsOmega} we plot the value of
the Josephson frequency as a function of the RF coupling. As mentioned in
the article this coupling dependence of the Josephson frequency
may impose a limit to the accuracy of the Josephson voltage-frequency relation.\\

{\centering\subsection*{Numerically calculated eigenstates}}

In general, solving analytically the eigenstates 
of the four state system of Figure 
\ref{fig:level_scheme}
produces extremely cumbersome formulas which do not give much insight to the problem. 
However, to illustrate our arguments, 
we give here numerically calculated eigenstates in certain selected cases. 
Since we are now solving the eigenstates of a multiple state system where 
the states are coupled and have energy differences (detunings) between 
them, this is similar to dressed state descriptions of some quantum optics 
systems \cite{cohen_tann}.\\

{\centering\subsubsection*{Zero detunings}}

Let us consider a system with $U/J=-5.0$.
In case of zero detunings $\delta_{13}/J=\delta_{24}/J=0$
but having finite couplings, here
$\Omega_{13}/J=\Omega_{24}/J=1$, the dressed states are the following 
\begin{align}
\ket{I}&= -0.6672\ket{12}+0.2341 \ket{14}-0.2341 \ket{23} -0.6672\ket{34},\\
\ket{II}&= 0.7071\ket{12} -0.7071 \ket{34},\\
\ket{III}&= 0.7071 \ket{14}+0.7071 \ket{23},\\
\ket{IV}&= -0.2341\ket{12}+0.6672 \ket{14}-0.6672 \ket{23}-0.2341\ket{34},\label{dress1}
\end{align}
with the eigenenergies
\begin{equation}
E_{I}=-5.7016, \quad E_{II}= -5.0, \quad E_{III}= 0.0, \quad E_{IV}= 0.7016. \label{dressE1}
\end{equation}
In general, if the initial state of the system is one of eigenstates 
of the Hamiltonian (a dressed state), there will be no dynamics. 
Here the state $\ket{II}$ is what one could consider in our small 
system as an analogue of the BCS state for two superfluids: 
a superposition of pairs. Having that state as the initial state, 
no dynamics takes place; this corresponds to the absence of Josephson 
oscillations for zero voltage (zero detuning) in case of identical 
superconductors. But if the phase difference between the 
superconductors, or their other properties, would be such that the 
initial state is no longer identical to one of the eigenstates 
(\textit{e.g.} $\ket{II}$), then there will be dynamics. That corresponds to 
the DC Josephson effect, caused by a phase difference between two 
superconductors instead of a voltage. This is just the standard 
two-mode description of Josephson physics found in many text-books 
(the two modes correspond to the pairs (superconductors) on both 
sides of the junction, here to the states $\ket{12}$ and $\ket{34}$).\\

{\centering\subsubsection*{Symmetric detunings}}

Now let us look at the effect of the detunings when they are symmetric. 
We take here the values $\delta_{13}/J=\delta_{24}/J=1$.
Then the eigenstates are
\begin{align}
\ket{I}&= 0.9581\ket{12} -0.1737\ket{14} +0.1737\ket{23}+0.1470\ket{34},\\
\ket{II}&= -0.2072\ket{12} -0.17 \ket{14} +0.17 \ket{23} +0.9483\ket{34},\\
\ket{III}&= 0.7071 \ket{14}+0.7071 \ket{23},\\
\ket{IV}&= 0.1976\ket{12}+0.664\ket{14} -0.664\ket{23} +0.2813\ket{34},\label{dress2}
\end{align}
with the eigenenergies
\begin{equation}
E_{I}=-6.3626, \quad E_{II}=-4.3586, \quad E_{III}= 0.0, \quad E_{IV}=0.7212. \label{dressE2}
\end{equation}
Now one can see that the states $\ket{12}$ and $\ket{34}$ do not appear 
any more symmetrically in the eigenstates. 
This leads to the Josephson oscillations for finite detuning (finite voltage), if we have an
initial state that contains only $\ket{12}$ and $\ket{34}$, as would be the analogue for
superconductors.

Note that the eigenstates contain unpaired states. These can then 
contribute to the dynamics. In this sense, the four state system 
description goes beyond the simple two-mode text-book description of the Josephson 
effect. Of course, these unpaired contributions are implicitly 
involved in more elaborate descriptions of Josephson physics, as 
is for instance evident from the fact that our self-consistent linear response 
calculation could reveal the asymmetry of the critical currents 
in the asymmetric detunings case. 

However, in previous literature, the unpaired contributions have 
not been paid attention to, because, as is clear from the symmetric 
structure of the above eigenstates with respect to states $\ket{14}$ and $\ket{23}$, 
the contributions of these states will be \textit{equal in magnitude} in the dynamics. 
Therefore it appears as if only pairs are tunneling and 
no single particles exist during the dynamics: 
but in fact, the single particles 
exist, it just happens that the expectation values of single 
particles in states $\ket{14}$ and $\ket{23}$ at any given time of the dynamics are the same.
Therefore it seems like $\ket{1}$ and $\ket{2}$ are tunneling together. 
Another reason why the single particle contributions have not 
been notified earlier is that it is known 
that the standard single particle currents require detunings 
(voltages) above twice the gap; how could they then participate 
in oscillations at the below gap Josephson frequency? In this manuscript, we have 
shown that it is \textit{the interference term of these single particle currents} 
(i.e. a beating of the standard single particle contributions) 
that contributes to the Josephson oscillations.\\      

{\centering\subsubsection*{Asymmetric detunings}}

Finally, let us choose different detunings with $\delta_{13}/J=1$ and $\delta_{24}/J=2$
\textit{i.e.} we consider the "different voltages spin up and spin down electrons" case. 
Then the eigenstates are
\begin{align}
\ket{I}&= 0.9704\ket{12} -0.1684\ket{14}+0.1454\ket{23}+0.0944\ket{34},\\
\ket{II}&= -0.1582\ket{12} -0.2291\ket{14} +0.1774\ket{23}+0.944\ket{34},\\
\ket{III}&= 0.0884\ket{12} +0.9138\ket{14} +0.3586\ket{23} +0.1692\ket{34},\\
\ket{IV}&= -0.1599\ket{12} -0.2902\ket{14} +0.9049\ket{23} -0.2672\ket{34},\label{dress3}
\end{align}
with the eigenenergies
\begin{equation}
E_{I}= -6.8234, \quad E_{II}=-3.9305, \quad E_{III}=-0.2182, \quad E_{IV}=0.9721. \label{dressE3}
\end{equation}
The observations of the dynamics in the previous case of symmetric detunings apply 
here as well, with the important difference that now the single 
particle contributions $\ket{14}$ and $\ket{23}$ do not appear symmetrically in 
the eigenstates. It is then understandable that asymmetry with respect to them appears 
also in the dynamics: this is the asymmetry in critical currents that we have 
predicted and explained. Note that by finding and explaining this 
asymmetry, we could provide insight also to the standard symmetric case as 
explained in the case of symmetric detunings: the Josephson effect consists of 
interference terms of pair and single particle tunneling processes, 
in particular, the interference of standard single particle currents contributes 
to the Josephson effect also in the symmetric case, although it 
appears as pair tunneling due to the symmetry. In the asymmetric case, the single 
particle interference term then clearly manifests itself, and the
novel intuitive understanding of the Josephson effect proposed by us, namely 
interferences in pair and single particle tunnelings, can be put under a direct 
experimental test.

%%%%%%%%%%%%%%%%%%%%%%%%%%%%%%%%%%%%%%%%%%%%%%%%%%%%%%%%%%%%%%%%%%%%%%%%%
%%%%%%%%%%%%%%%%%%%%%%%%%%%%%%%%%%%%%%%%%%%%%%%%%%%%%%%%%%%%%%%%%%%%%%%%%
%%%%%%%%%%%%%%%%%%%%%%%%%%%%%%%%%%%%%%%%%%%%%%%%%%%%%%%%%%%%%%%%%%%%%%%%%
%%%%%%%%%%%%%%%%%%%%%%%%%%%%%%%%%%%%%%%%%%%%%%%%%%%%%%%%%%%%%%%%%%%%%%%%%